\documentclass[aps,prd,superscriptaddress,preprint,12pt,nofootinbib]{revtex4} 
\pdfoutput=1

\usepackage{amsmath,amsfonts,amssymb,graphicx,natbib,bm}
\usepackage{color}
\usepackage{slashed}    
\usepackage{hyperref}
\usepackage{breakurl}
\usepackage{multirow}

\begin{document}

\preprint{ANL-HEP-PR-13-39}
\preprint{EFI-13-30}

\title{Constraints on a Very Light Sbottom}

\author{Brian Batell}
\affiliation{Enrico Fermi Institute and Department of Physics, University of Chicago, Chicago, IL 60637}

\author{Carlos E. M. Wagner}
\affiliation{Enrico Fermi Institute and Department of Physics, University of Chicago, Chicago, IL 60637}
\affiliation{HEP Division, Argonne National Laboratory, 9700 Cass Ave., Argonne, IL 60439}
\affiliation{Kavli Institute for Cosmological Physics, University of Chicago, Chicago, IL 60637}

\author{Lian-Tao Wang}
\affiliation{Enrico Fermi Institute and Department of Physics, University of Chicago, Chicago, IL 60637}
\affiliation{Kavli Institute for Cosmological Physics, University of Chicago, Chicago, IL 60637}

%\pagestyle{empty}
%\date{\today}

%%%%%%%%%%%
\begin{abstract}
We investigate the phenomenological viability of a very light bottom squark, with a mass less than half of the $Z$ boson mass.
The decays of the $Z$ and Higgs bosons to light sbottom pairs are, in a fairly model independent manner, strongly constrained by the precision electroweak data and Higgs signal strength measurements, respectively. These constraints are complementary to direct collider searches, which depend in detail on assumptions regarding the superpartner spectrum and decays of the sbottom. In particular, if the lightest sbottom has a mass below about 15 GeV,  compatibility with these measurements is possible only in a special region of parameter space in which the couplings of the lightest sbottom to the $Z$ and Higgs are suppressed. In this region, the second sbottom is predicted to be lighter than about 300 GeV and can also be searched for directly at the LHC. We also survey relevant collider searches for canonical scenarios with a bino, gravitino, or singlino LSP in the compressed and stealth kinematic regimes and provide suggestions to cover remaining open regions of parameter space. 
\end{abstract}
%%%%%%%%%%%%%%
% =============================================================================
\maketitle
\newpage

%%%%%%%%%%%%%%%%%%%%%%%%%%%%%%
%%%%%%%%%%%%%%%%%%%%%%%%%%%%%%
\section{Introduction}\label{sec:intro}

With the discovery of the Higgs boson~\cite{ATLAS-Higgs, CMS-Higgs} and a vibrant experimental program to measure its couplings and quantum numbers, significant progress is being made in our understanding of the physics of electroweak symmetry breaking. 
However, the existence of a Higgs particle with Standard Model (SM)-like properties accentuates the hierarchy problem. Weak scale supersymmetry (SUSY) provides one of the most compelling solutions to the hierarchy problem, and it is intriguing that the Higgs boson mass, $m_{h^0} \sim 126$ GeV, lies within the range predicted by the Minimal Supersymmetric Standard Model (MSSM)~\cite{Okada:1990vk,Ellis:1990nz,Haber:1990aw,Casas:1994us,Carena:1995bx,Carena:1995wu,Haber:1996fp,Heinemeyer:1998yj,Heinemeyer:1998np,Carena:2000dp,Martin:2002wn,Degrassi:2002fi}.

However, with the completion of the 8 TeV run at the LHC, new strong limits have been placed on a variety of SUSY scenarios and spectra. For example, searches based on several jets in association with large missing transverse energy have lead to bounds on squarks and gluinos in the TeV range~\cite{ATLAS-CONF-2013-047,CMS-PAS-SUS-13-012}.  While we continue pursuing even heavier superpartners in canonical SUSY models, it is critical that we do not overlook possible loopholes in the experimental searches that may be present in non-standard scenarios. 

One interesting example in this regard is a very light bottom squark with a mass well below the kinematic LEP bound of 100 GeV. 
Such a light sbottom has been considered on numerous occasions in the past, including studies of its effects on the precision electroweak data~\cite{Carena:2000ka}, 
the Tevatron measurement of the bottom quark cross section~\cite{Berger:2000mp}, new Higgs boson decay channels~\cite{Berger:2002vs}, 
and myriad related phenomenology~\cite{Berger:2001jb,Berger:2002gu,Cao:2001rz,Grossman:2002bu,Cho:2002mt,Becher:2002ue,Cheung:2002rk,Chiang:2002wi,Luo:2003uw,Janot:2004cy}. 
More recently, it was observed that a light sbottom can mediate large spin-independent dark matter scattering cross sections which may be relevant for some of the anomalies in the direct detections experiments~\cite{Arbey:2012bp,Arbey:2012na,Gondolo:2013wwa,Arbey:2013aba}. 
Regardless of any particular phenomenological motivation, it is of general interest to understand the constraints on this scenario and the allowed regions of parameter space. 

The focus of the present work is an investigation of the constraints on the sbottom parameter space implied by the precision electroweak and Higgs signal strength data. As we will discuss in detail below, these constraints only weakly depend on the assumed decay mode of the lightest sbottom. Therefore, conservative, robust statements can be made with regard to the allowed parameter space. This is in contrast with limits from direct searches at colliders, which strongly depend on the assumptions regarding the superpartner spectum and sbottom decay channels. We find that the combination of electroweak and Higgs data restrict the parameter space to a special region in which the sbottom couplings to the $Z$ and Higgs boson are suppressed. Furthermore, these constraints imply that the second sbottom should be lighter than about 300 GeV if the lightest sbottom has a mass of 15 GeV or less. Concerning the precision electroweak data, we stress the importance of the measurement of the total hadronic cross section at the $Z$ pole, 
$\sigma_{\rm had}^0$, which, independent of any particular model, places a strong constraint on any new decay modes of the $Z$ boson. 

In light of these model independent constraints, we provide a survey of the existing collider constraints on a canonical scenario in which the LSP is a neutral fermion, such as a bino, gravitino, or singlino. We particularly emphasize the importance of the direct searches for the second sbottom.  Searches are suggested to cover the remaining holes in the sbottom parameter space. 

The outline of the paper is as follows: 
We begin in Sec.~\ref{sec:LightSbottom} with a description of the sbottom sector and the radiative corrections to the bottom Yukawa coupling. 
In Sec.~\ref{sec:EWPD} and Sec.~\ref{sec:higgs} we investigate the constraints from the precision electroweak and the Higgs signal strength data, respectively, on the sbottom parameter space. In Sec.~\ref{sec:stops} we discuss the effects of the stops on the $\rho$ parameter and Higgs mass. 
We then combine these model independent constraints in Sec.~\ref{sec:combined}. In Sec.~\ref{sec:collider} we discuss the collider constraints on light sbottoms.
Finally, in Sec.~\ref{sec:conclusions} we present our conclusions.

%%%%%%%%%%%%%%%%%%%%%%%%%%%%%%
%%%%%%%%%%%%%%%%%%%%%%%%%%%%%%
\section{Sbottom Sector}\label{sec:LightSbottom}

We begin by describing our conventions for the sbottom sector. 
In terms of the gauge eigenstates $(\tilde b_L, ~\tilde b_R)$, the sbottom mass matrix is given at tree level by
\begin{equation}
\left( 
\begin{array}{cc}
m_{Q_3}^2 + m_b^2 + D_L
& m_b (A_b -\mu \tan\beta)\\
m_b (A_b -\mu \tan\beta)   &  m_{D_3}^2 + m_b^2 + D_R
\end{array}
\right), 
\label{eq:sbottom-mass}
\end{equation}
where $m^2_{Q_3}$, $m^2_{D_3}$ are the left- and right-handed squark soft mass parameters, $A_b$ is the soft trilinear coupling, $\mu$ is the supersymmetric Higgs mass parameter, $\tan \beta$ is the ratio of up and down type Higgs vacuum expectation values, and
$D_L =   m_Z^2 \cos{2 \beta} \left( -\tfrac{1}{2}  +\tfrac{1}{3} s_W^2 \right)$, 
$D_R  =    m_Z^2 \cos{2 \beta} \left(  - \tfrac{1}{3} s_W^2 \right)$.
For simplicity, we will assume all parameters are real.  The physical sbottom mass eigenstates are related to the gauge eigenstates through the orthogonal transformation:
\begin{equation}
\label{stopX}
\left(
\begin{array}{c}
\tilde b_L \\
\tilde b_R
\end{array}
 \right)
 = 
\left(
\begin{array}{cc}
 \sin \theta_{b} & \cos\theta_{b} \\
\cos \theta_{b} &  -\sin \theta_{b}
\end{array}
 \right)
\left(
\begin{array}{c}
\tilde b_1 \\
\tilde b_2
\end{array}
 \right),
\end{equation}
where the mixing angle $\theta_b$ satisfies
\begin{equation}
\tan 2 \theta_{b} = \frac{ 2 m_b (A_b -\mu \tan \beta )}{m_{D_3}^2- m_{Q_3}^2 + D_R - D_L},
\end{equation}
and lies in the range $[-\pi/2, \pi/2]$.

\subsection{\bf Radiative corrections to $y_b$}

The bottom Yukawa coupling $y_b$ can receive substantial corrections at one loop \cite{Pierce:1996zz,Carena:1998gk,Carena:2002es,Guasch:2001wv}, which are important for the light sbottom regime considered in this work. The correction can be written in terms of the parameter $\Delta_b$, such that
\begin{equation}
y_b \rightarrow \frac{m_b}{v \cos\beta (1+\Delta_b)}.
\end{equation}
This corrects the sbottom mass matrix in Eq.~(\ref{eq:sbottom-mass}). The dominant effect occurs in the off-diagonal term, which becomes $m_b (A_b - \mu\tan\beta)/(1+\Delta_b)$. 
Rather than compute $\Delta_b$ directly, we will instead find it useful to define the effective parameters, 
\begin{equation}
\label{eq:effective parameters}
A_{b,{\rm eff}} = \frac{A_b	}{1+\Delta_b}, ~~~~~~~~~\tan\beta_{b,{\rm eff}} = \frac{\tan\beta	}{1+\Delta_b}. 
\end{equation}
In this way, one can absorb the radiative corrections to the sbottom masses into a redefinition of $A_b$ and $\tan\beta$ and use the tree-level equations for the masses and mixing angles in the sbottom sector. We note that a complete removal of the $\Delta _b$ dependence is not possible, since it appears also in the diagonal mass entries in~(\ref{eq:sbottom-mass}). 
However, the term proportional to $m_b^2$ is numerically small and the $D$-term contribution is proportional to $\cos 2\beta \approx -1 +2/\tan^2\beta$ which is approximately constant for $\tan\beta \gg 1$, so that we can safely use Eq.~(\ref{eq:effective parameters}) in the regime of interest.

One may further worry that a mismatch can occur in the couplings of the sbottom to other fields. For our purposes, the couplings of the Higgs to the sbottoms will be important when we examine the constraints imposed by the Higgs signal strength data. In particular, the coupling $h^0-\tilde b_1 - \tilde b_1^*$, 
which is given in the Appendix in Eq.~(\ref{eq:sfermion-higgs}) contains the same factor $m_b(A_b - \mu \tan\beta)$, that appears in the off-diagonal entry of the sbottom mass matrix, and thus the $\Delta_b$ correction can be absorbed in this coupling using Eq.~(\ref{eq:effective parameters}). Furthermore, the stop mass matrix and couplings to the Higgs are independent of $\tan\beta$ in the large $\tan\beta$ limit. Therefore, we can safely use the effective parameters to absorb the radiative corrections to the bottom Yukawa coupling. For the remainder of this paper we will therefore use the tree level formulae for the sbottom masses, mixing angles, and couplings, substituting the effective parameters
defined above in~(\ref{eq:effective parameters}) and neglecting small corrections to this approximation which are of order $y_b^2$ and $1/\tan\beta$. We will also drop everywhere the ``eff" subscript, although the use of the effective parameters~(\ref{eq:effective parameters}) is implied.
As we will see, this approach allows for an analytical comprehension of the predictions and constraints on the parameters of the model, which is more difficult to obtain with a parameter space scan.

%%%%%%%%%%%%%%%%%%%%%%%%%%%%%%
%%%%%%%%%%%%%%%%%%%%%%%%%%%%%%
\section{Precision Electroweak Data}\label{sec:EWPD}

In this section we study the impact of sbottoms lighter than $m_Z/2$ on precision electroweak measurements. 
Such sbottoms can be produced in the  decays of $Z$ bosons via $Z\rightarrow \tilde b_1 \tilde b_1^*$ and through the continuum reaction $e^+ e^- \rightarrow \gamma^*, Z^* \rightarrow  \tilde b_1 \tilde b_1^*$ and will therefore enter into the predictions for a variety of precision observables. 

However, in order to make sharp predictions, it is necessary to understand how the sbottom decay products are ``counted" -- that is, whether the sbottom events populate the signal regions defined in the various analyses underlying the measurements of the precision observables. This is a difficult question on two fronts. First, the reconstruction of the sbottom is model dependent and requires assumptions about the decay channel of the sbottom and masses of the sbottom and its decay products. 
Second, the LEP experiments employed a complex and evolving set of strategies to perform these measurements. 
For instance, for the measurements of $R_b$ and $A_{FB}^b$ at ALEPH, several algorithms were developed to effectively identify $b$ quarks, including the presence of a high momentum lepton (presumed to originate from a $B$-hadron)~\cite{Decamp:1990cp,Decamp:1991eg}, a lifetime tag based on longevity of the $B$-hadron~\cite{Buskulic:1993ka}, and multivariate analysis of event shape variables~\cite{Buskulic:1993kb}. It is likely that certain sbottom decays would populate these signal regions, but to properly address this issue would require a detailed simulation of these algorithms that goes beyond the scope of this paper. 

To simplify the analysis, we will make a reasonable assumption about how the sbottom is ``counted'' by the experiments at LEP and SLC. 
We can conceive of the following possibilities:
\begin{enumerate}
\item {\it ``Invisible'' sbottom}: {\it i.e.}, the sbottom events do not populate the signal regions in searches entering into the precision measurements. 
This may happen, {\it e.g.}, if the sbottom decays via $\tilde b_1 \rightarrow b + \tilde \chi^0$ in the compressed kinematic regime,  
$m_{\tilde b_1} \gtrsim m_{\tilde \chi^0} \gg m_b$, such that the $b$ quark is too soft to pass basic event selection cuts for hadronic $Z$ decays~\cite{Decamp:1989tf}.  
\item {\it  Hadronic sbottom}: This can occur, for instance, if the sbottom decays through an R-parity violating $UDD$ coupling to jets. 
\item {\it Sbottom counted as a $b$ quark}. This will happen, for example, in the stealth kinematic regime, $m_{\tilde b_1} \gtrsim m_b \gg m_{\tilde \chi^0}$
and the sbottom decays via $\tilde b_1 \rightarrow b + \tilde \chi^0$. 
\end{enumerate}
In the next subsection, Sec.~\ref{subsec:sbottom-EWPD}, we will describe the predictions for the precision electroweak observables for the three sbottom reconstruction scenarios listed above. Following this we will present a quantitative analysis of the constraints obtained through a global fit to the precision data.
In particular, we will see that for very light sbottoms with masses below about 15 GeV, the lightest sbottom must be largely decoupled from the $Z$ boson in order to provide a good description of the data. This conclusion is conservative and robust, {\it i.e.}, it does not depend on which hypothesis is made regarding the sbottom reconstruction.

\subsection{Sbottom contributions to precision observables}\label{subsec:sbottom-EWPD}

%%%%%%%%%%%%%%%%%%%%%%%%%%%%
\subsubsection{``Invisible" sbottom} 

If the lightest sbottom is ``invisible", {\it i.e.}, it does not populate any of the visible search channels that enter in the precision measurements, it will still affect the prediction for the total $Z$ boson width, which is exquisitely measured from the $Z$ lineshape~\cite{LEPEWWG}. A light sbottom leads to the new decay channel,
\begin{equation}
Z \rightarrow \tilde b_1 \tilde b_1^*,
\label{ZdecaySbottom}
\end{equation}
which is kinematically allowed provided $m_{\tilde b_1} < m_Z /2$.  The partial decay width for the process (\ref{ZdecaySbottom}), 
$\Gamma_{\tilde b \tilde b^*} \equiv \Gamma(Z\rightarrow \tilde b_1 \tilde b^*_1)$,
 is given by
\begin{equation}
\Gamma_{\tilde b \tilde b^*}
 = g_{Z \tilde b_1 \tilde b_1^*}^2    \frac{ \alpha  \,  m_Z}{4 s_W^2 c_W^2} \left( 1 - \frac{ 4 m_{\tilde b_1}^2  }{m_Z^2}  \right)^{3/2}, 
 \label{eq:Z-Sbottom-decay}
\end{equation}
where the coupling $g_{Z \tilde b_1 \tilde b_1^*}$ is defined as
\begin{eqnarray}
g_{Z \tilde b_1 \tilde b_1^*}  
 &  =   &    - \frac{1}{2} \sin^2{ \theta_b} +  \frac{1}{3} s_W^2 .
\label{eq:Z-sbottom-coupling}
\end{eqnarray}
Note that this coupling vanishes for a mixing angle $\theta_b \approx \pm 0.4$. Thus, for mixing angles near this decoupling region, the new decay mode (\ref{ZdecaySbottom}) is suppressed and the predictions for the precision observables are close to their SM values. Away from the decoupling region, however, there will be dramatic departures from the SM predictions due to the new decay mode (\ref{ZdecaySbottom}).

As an example, a light purely right-handed sbottom ($m_{\tilde b_1} = 15\, {\rm GeV}  \ll m_Z$, $\theta_b = 0$) yields a contribution to the total width of  
$\Gamma_{\tilde b \tilde b^*} \approx 5$ MeV. For comparison, the measured value is $(\Gamma_Z)_{\rm exp} = 2.4952\pm0.0023$ GeV~\cite{LEPEWWG}, 
while the SM prediction from the Gfitter group $(\Gamma_Z)_{\rm SM} = 2.4954 \pm 0.0014$ GeV~\cite{Baak:2012kk}. 
Thus,  a pure $\tilde b_R$ is marginally allowed by this measurement alone. 

In fact, a more constraining measurement in this scenario is the total hadronic cross section at the $Z$ pole, $\sigma_{\rm had}^0$. The prediction for this observable is also modified by the presence of light sbottoms, since the cross section on the resonance depends on the total $Z$ boson width. 
The prediction for this observable is given by
\begin{eqnarray}
\sigma_{\rm had}^0 & = & \frac{12 \pi}{m_Z^2} \frac{\Gamma_{e^+ e^-}  \Gamma_{\rm had}}{\Gamma_Z^2} \nonumber \\
& = & (\sigma_{\rm had}^0)_{\rm SM}\left(1+\frac{\Gamma_{\tilde b \tilde b^*} }{(\Gamma_Z)_{\rm SM}}\right)^{-2}  \label{sigmahad-inv}\\
& \approx & (\sigma_{\rm had}^0)_{\rm SM}\left(1-2 \frac{\Gamma_{\tilde b \tilde b^*} }{(\Gamma_Z)_{\rm SM}}\right), \nonumber
\end{eqnarray}
where $\Gamma_{e^+ e^-} \equiv \Gamma(Z\rightarrow e^+ e^-) $,  $\Gamma_{\rm had}$ is the total hadronic width of the $Z$, 
and in the last step we have used $\Gamma_{\tilde b \tilde b^*}  \ll  (\Gamma_Z)_{\rm SM}$. 
The measured value and SM prediction for this observable are~\cite{LEPEWWG},~\cite{Baak:2012kk}:
\begin{eqnarray}
(\sigma_{\rm had}^0)_{\rm exp} & = & 41.540 \pm 0.037~{\rm nb}, \nonumber \\ 
(\sigma_{\rm had}^0)_{\rm SM} & = & 41.479 \pm 0.014~{\rm nb}, 
\end{eqnarray}
which disagree at the $1.5\sigma$ level. One observes from Eq.~(\ref{sigmahad-inv}) that the presence of the light sbottom only serves to increase the tension as it strictly decreases the theory prediction. For example, a 15 GeV purely right-handed sbottom leads to a prediction $(\sigma_{\rm had}^0)  =  41.313$ nb, which is at odds with the measured value of the $5-6\sigma$ level and is therefore disfavored. 
We note that this observable was not considered in the recent paper~\cite{Arbey:2013aba} invoking a light sbottom to mediate a large neutralino direct detection cross section, and strongly constrains their scenario. \\ 

We pause here to emphasize that, independent of any particular model, that the total hadronic cross section $\sigma_{\rm had}^0$ yields a stronger constraint on a new contribution to the $Z$ boson invisible width. If we take the data and SM theory predictions at face value, then demanding that these observables agree to within $2\sigma$ we find that $\delta\Gamma_{Z,{\rm inv}} < 0.5$ (5.6) MeV from $\sigma_{\rm had}^0$ ($\Gamma_Z$). Alternatively, one might be skeptical about the $1.5$ sigma discrepancy in  
$\sigma_{\rm had}^0$.  Even in this case, if we demand that the new physics contribution to these observables is less than 2 times the combined experimental and theoretical uncertainty, then we obtain $\delta\Gamma_{Z,{\rm inv}} < 2.4$ (5.4) MeV.

\subsubsection{Hadronic sbottom} 

If the sbottom decays to hadrons then one expects that it will contaminate the hadronic width of the $Z$ boson, $\Gamma_{\rm had}$. The hadronic width enters into a number of precision observables, including $\Gamma_Z$ and $\sigma_{\rm had}^0$ already discussed above, as well as $R_\ell \equiv \Gamma_{\rm had}/\Gamma_\ell$, with $\Gamma_\ell$ the leptonic width of the $Z$, $R_b \equiv \Gamma_{b}/\Gamma_{\rm had}$ and $R_c$ (defined analogously). As in the case of ``invisible" sbottom decays discussed above, the total peak hadronic cross section measurement is particularly constraining. The prediction for $\sigma_{\rm had}^0$ in this case is
\begin{eqnarray}
\sigma_{\rm had}^0 & = & (\sigma_{\rm had}^0)_{\rm SM}\left(1+\frac{\Gamma_{\tilde b \tilde b^*} }{(\Gamma_{\rm had})_{\rm SM}}\right)\left(1+\frac{\Gamma_{\tilde b \tilde b^*} }{(\Gamma_Z)_{\rm SM}}\right)^{-2} \\
& \approx &  (\sigma_{\rm had}^0)_{\rm SM}  
\left(1-0.57 \frac{\Gamma_{\tilde b \tilde b^*} }{(\Gamma_Z)_{\rm SM}}\right),
\nonumber
\end{eqnarray} 
where in the second line we have assumed $\Gamma_{\tilde b \tilde b^*} \ll  (\Gamma_{\rm had})_{\rm SM} < (\Gamma_Z)_{\rm SM}$ and have used the SM predictions for the hadronic and total widths. 
Note that the hadronic sbottom leads to a lower value of $\sigma_{\rm had}^0$ than in the SM, but the suppression is not as severe as in the case of the invisible sbottom (see 
Eq.~(\ref{sigmahad-inv})). For example, a 15 GeV purely right-handed sbottom leads to a prediction $(\sigma_{\rm had}^0)  =  41.432$ nb, which is lower than the experimental value by about $ 3 \sigma$. 

\subsubsection{$\tilde b_1$ is counted as $b$} 

Finally, it may happen in some scenarios that the sbottom mimics a $b$ quark through its decay. 
In addition to the effects described in the hadronic sbottom case above, there are a few additional observables which are affected in the case that a sbottom is reconstructed as a $b$ quark, as we now discuss. 

The first observable to consider is the forward-backward asymmetry of the bottom quark $A_{FB}^b$. In general the forward-backward asymmetry is defined as
\begin{equation}
A_{FB}  = \frac{ \sigma_F - \sigma_B}{  \sigma_F + \sigma_B },
\label{eq:AFB}
\end{equation} with 
$\sigma_{F,B}  = \pm \int_{0}^{\pm 1}  \!  d\!\cos\theta ( d\sigma / d\!\cos\theta)$. 
If the sbottom is counted as a $b$ quark, then the reaction $e^+ e^- \rightarrow \tilde b_1 \tilde b_1^*$ will contribute to this asymmetry. 
Since the sbottom is a scalar, the forward and backward cross sections for sbottom pair production are identical, and thus sbottom pair production will only contribute to the total cross section for $b\bar b$ production. In other words, the effect of the sbottom is to strictly increase the denominator in Eq.~(\ref{eq:AFB}), thus lowering the prediction for $A_{FB}^b$ with respect to the SM prediction. 

This is quite intriguing given the longstanding $2.4\sigma$ discrepancy in the measured and predicted values of $A_{FB}^b$, which are respectively given by~\cite{LEPEWWG},~\cite{Baak:2012kk}:
\begin{eqnarray}
\label{eq:AFBb-exp}
(A_{FB}^b)_{\rm exp} &  =  & 0.0992 \pm 0.0016, \\
\label{eq:AFBb-SM}
(A_{FB}^b)_{\rm SM}  & =  &  0.1032^{+0.0004}_{-0.0006} .
\end{eqnarray}
Since the SM prediction (\ref{eq:AFBb-SM}) is larger than the measured value (\ref{eq:AFBb-exp}), a small sbottom pair production cross section will improve the agreement between theory and data. 

It follows from Eq.~(\ref{eq:AFB}) that the prediction for $A_{FB}^b$ is 
\begin{equation}
A_{FB}^b = (A_{FB}^b)_{\rm SM} \left( 1+ \frac{ \sigma_{\tilde b \tilde b^*} }{ \sigma_{b \bar b} }\right)^{-1}.
\label{eq:AFBb-1}
\end{equation}
Here, $\sigma_{b \bar b}$ $(\sigma_{\tilde b \tilde b^*})$ 
are the total bottom (sbottom) production cross sections on the $Z$-pole, given by
\begin{eqnarray}
\label{eq:XS-bottom}
\sigma_{b \bar b} & \simeq & \frac{\pi \alpha^2 }{ s_W^4 c_W^4 \Gamma_Z^2}
 \left[ (g_{Le}^2 + g_{Re}^2   ) (g_{Lb}^2+g_{Rb}^2)\right] \left[ 1-\frac{m_b^2}{m_Z^2}\left(1 -\frac{6 g_{Lb} g_{Rb} }{g_{Lb}^2+g_{Rb^2}} \right)  \right]\left( 1-\frac{4 m_b^2}{m_Z^2}  \right)^{1/2}, ~~~~~~~ \\
 \label{eq:XS-sbottom}
 \sigma_{\tilde b \tilde b^*} & \simeq & \frac{\pi \alpha^2 }{ s_W^4 c_W^4 \Gamma_Z^2}
 \left[\frac{1}{2} (g_{Le}^2 + g_{Re}^2   ) (g_{Z \tilde b_1 \tilde b_1^*}^2)\right]\left( 1-\frac{4 m_{\tilde b_1}^2}{m_Z^2}  \right)^{3/2}, 
\end{eqnarray}
where 
$g_{Lf} = T^3_f - Q_f s_W^2$, $g_{Rf} = - Q_f s_W^2 $ for the fermion $f$, and  $g_{Z \tilde b_1 \tilde b_1^*}$ is the coupling of the $Z$ boson to the lightest sbottom mass eigenstate given in Eq.~(\ref{eq:Z-sbottom-coupling}). 
As numerical examples, we find that for $m_{\tilde  b_1} = 5.5$ GeV, a purely right-handed sbottom ($\theta_b = 0$) eases the tension in $A_{FB}^b$ to the $1.5\sigma$ level, while for a mixing angle $\theta \sim 0.7$, the discrepancy disappears completely. Unfortunately, in both cases there are other observables which are in tension with the measured values due to the effects of the sbottoms, as we will see below. 

Another observable that is affected in this scenario is $R_b$, the ratio of the $Z\rightarrow b \bar b$ partial width to the total hadronic width.
The prediction is given by 
\begin{equation}
R_b =  \frac{  \Gamma_{b \bar b} + \Gamma_{\tilde b \tilde b^*}  }{ 2 (\Gamma_{u \bar u}   + \Gamma_{d \bar d}) + \Gamma_{b \bar b} + \Gamma_{\tilde b \tilde b^*}  },
 \label{eq:Rb}
\end{equation} 
where $\Gamma_{\tilde b \tilde b^*}$ is given in Eq.~\ref{eq:Z-Sbottom-decay}.
As of September 2013, the predictions and measured values for $R_b$ are~\cite{Freitas:2012sy}
\begin{eqnarray}
\label{eq:Rb-exp}
(R_b)_{\rm exp} &  =  & 0.21629 \pm 0.00066, \\
\label{eq:Rb-SM}
(R_b)_{\rm SM}  & =  &0.21550\pm0.00003, 
\end{eqnarray}
which are consistent at the 1.2$\sigma$ level.  From Eq.~(\ref{eq:Rb}) above it is clear that the sbottom contribution can only lead to a larger prediction for $R_b$ than the SM, which if small, will improve the agreement even further.

\subsection{Global fit to the precision data}

We now investigate quantitatively the constraints from the precision electroweak data on a light sbottom under each of the possible sbottom reconstruction hypothesis presented at the beginning of Sec.~\ref{sec:EWPD}.  Our results are based on a global fit to the precision electroweak data, which closely follows the fit of the Gfitter group~\cite{Baak:2012kk}. 
There are 19 observables entering into the fit and 6 fit parameters (five SM parameters plus the sbottom mixing angle $\theta_b$) leading for 13 degrees-of-freedom (d.o.f.). 
We will investigate four different sbottom masses $m_{\tilde b_1}$ = (5.5, 15, 25, 35) GeV. 
A detailed description of the experimental observables and SM theory predictions that enter into the fit can be found in Ref.~\cite{Batell:2012ca}. 
For the SM theory predictions for $\Gamma_Z$ and $\sigma_{\rm had}^0$ we employ the recent numerical parameterizations presented in Ref.~\cite{Freitas:2013dpa} which include the complete two-loop electroweak corrections.

An important issue is the treatment of $\alpha_s(m_Z)$ in the fit. The standard procedure, which is followed by the LEP Electroweak Working Group~\cite{LEPEWWG}, the Particle Data Group (PDG)~\cite{Beringer:1900zz}, and Gfitter~\cite{Baak:2012kk} is to allow $\alpha_s$ to float in the fit, thereby providing an independent determination of this parameter. 
However, in the light sbottom scenario there is typically a strong preference for low values of $\alpha_s$, which raises the SM prediction for $\sigma_{\rm had}^0$ towards the measured value. 
 Due to this strong pull towards low values, we constrain $\alpha_s$ in our fit as we now discuss.
 
Besides the precision electroweak data, there are a number of other measurements of $\alpha_s$, including determinations from tau decays, lattice QCD, deep inelastic scattering (DIS), heavy quarkonia, and hadronic event shapes; see the PDG~\cite{Beringer:1900zz} and the recent review~\cite{Bethke:2012jm} for more details. The most recent world average, which includes a subset of the various $\alpha_s$ measurements, is $\alpha_s = 0.1184\pm0.0007$~\cite{Bethke:2012jm}. 
However, the central value of $\alpha_s$ must take into account the light sbottom contribution to its running, which is relevant in the extraploation of the dominant low energy determinations of $\alpha_s$  to the scale $m_Z$. Running from the scale $Q = (5.5, 15, 25, 35)$ GeV to $Q = m_Z$, we find that the sbottom induces an upward shift $\Delta \alpha_s \sim (0.0009, 0.0008,0.007,0.0006)$. In our fits we add this shift to the central value, and constrain $\alpha_s$ in our fit with the quoted uncertainty $0.0007$.

\subsubsection{Results}

\begin{figure*}
\begin{center}
\includegraphics[width=0.45\textwidth]{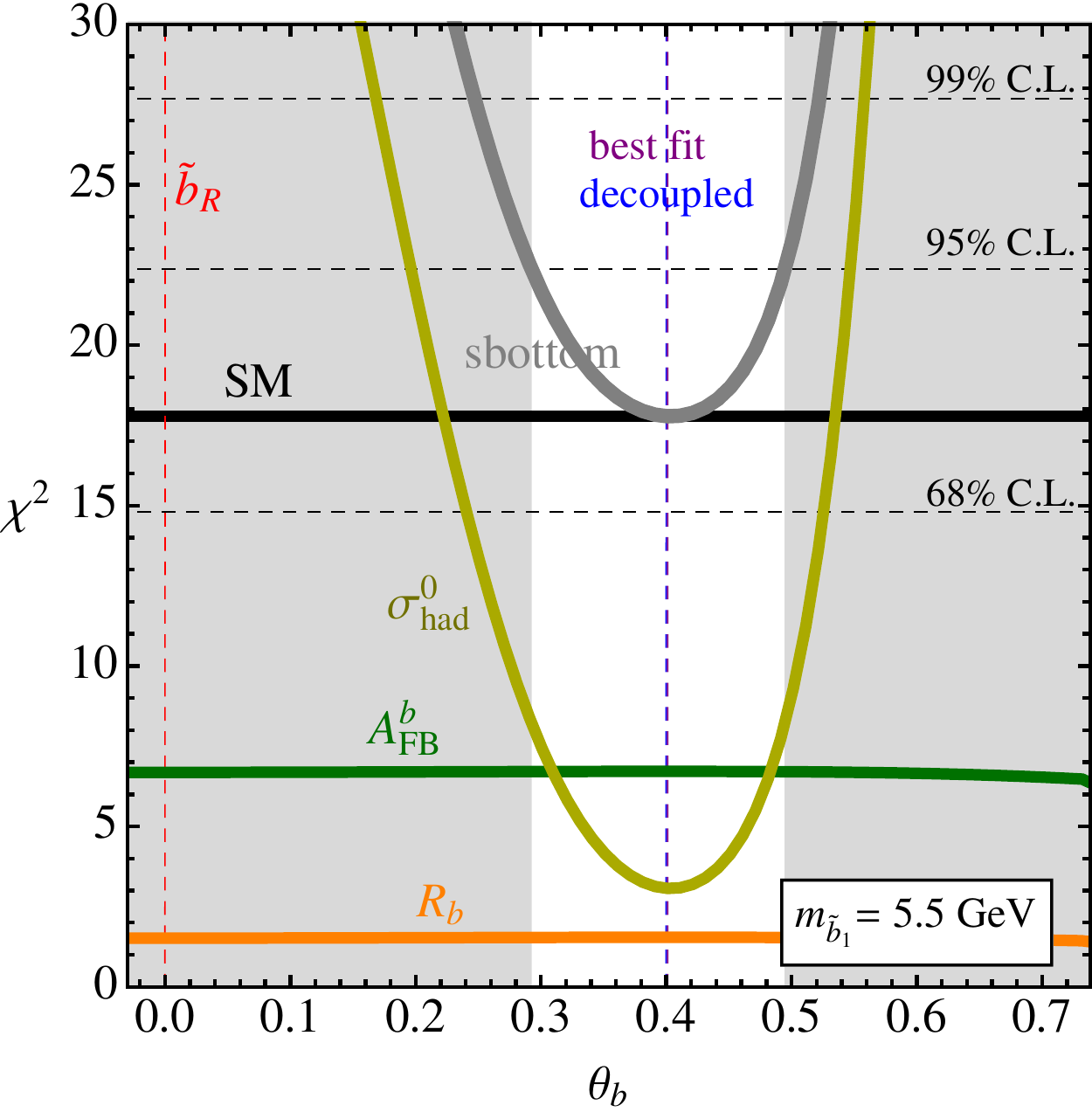}~~~
\includegraphics[width=0.45\textwidth]{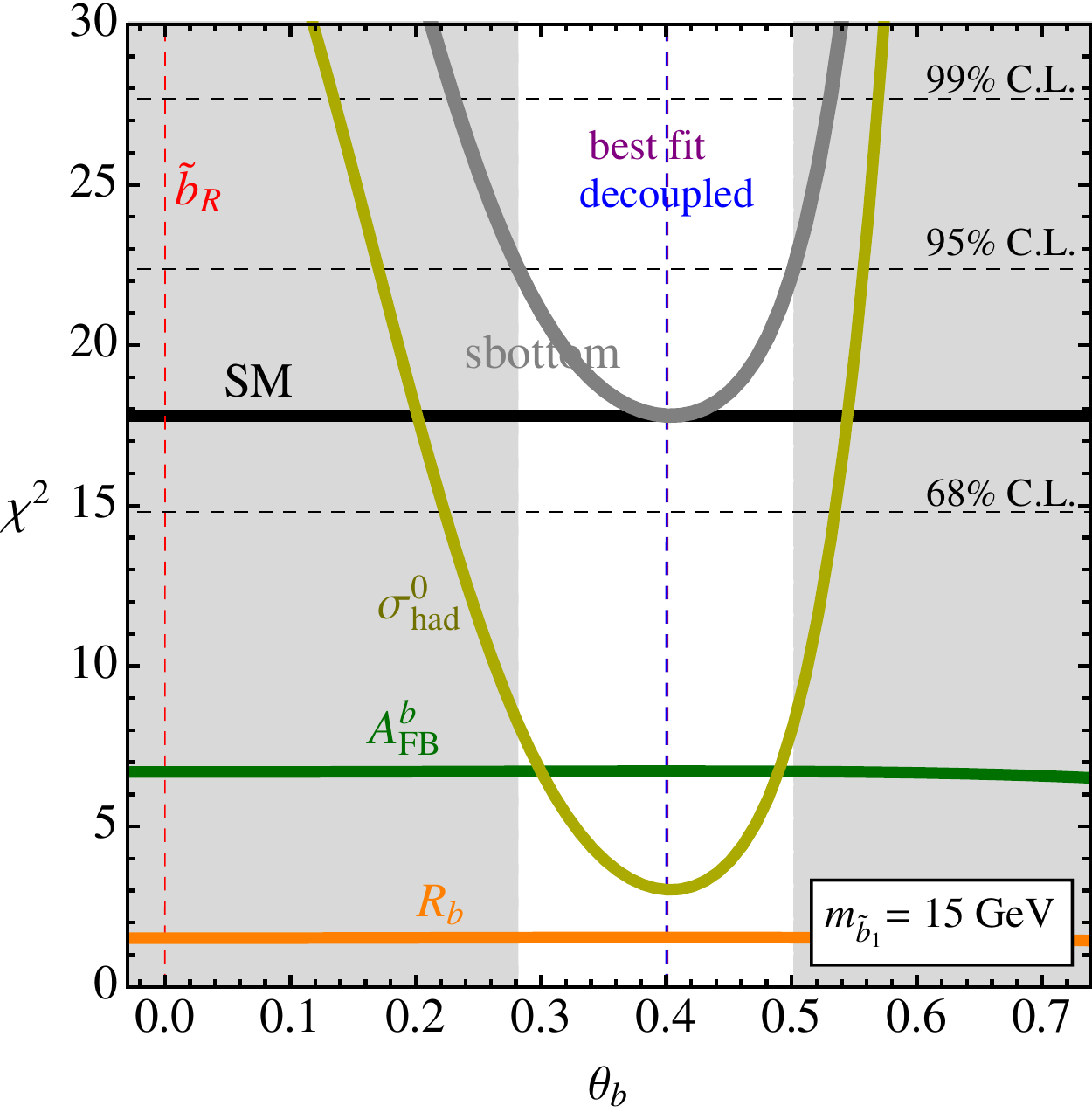} \\
\includegraphics[width=0.45\textwidth]{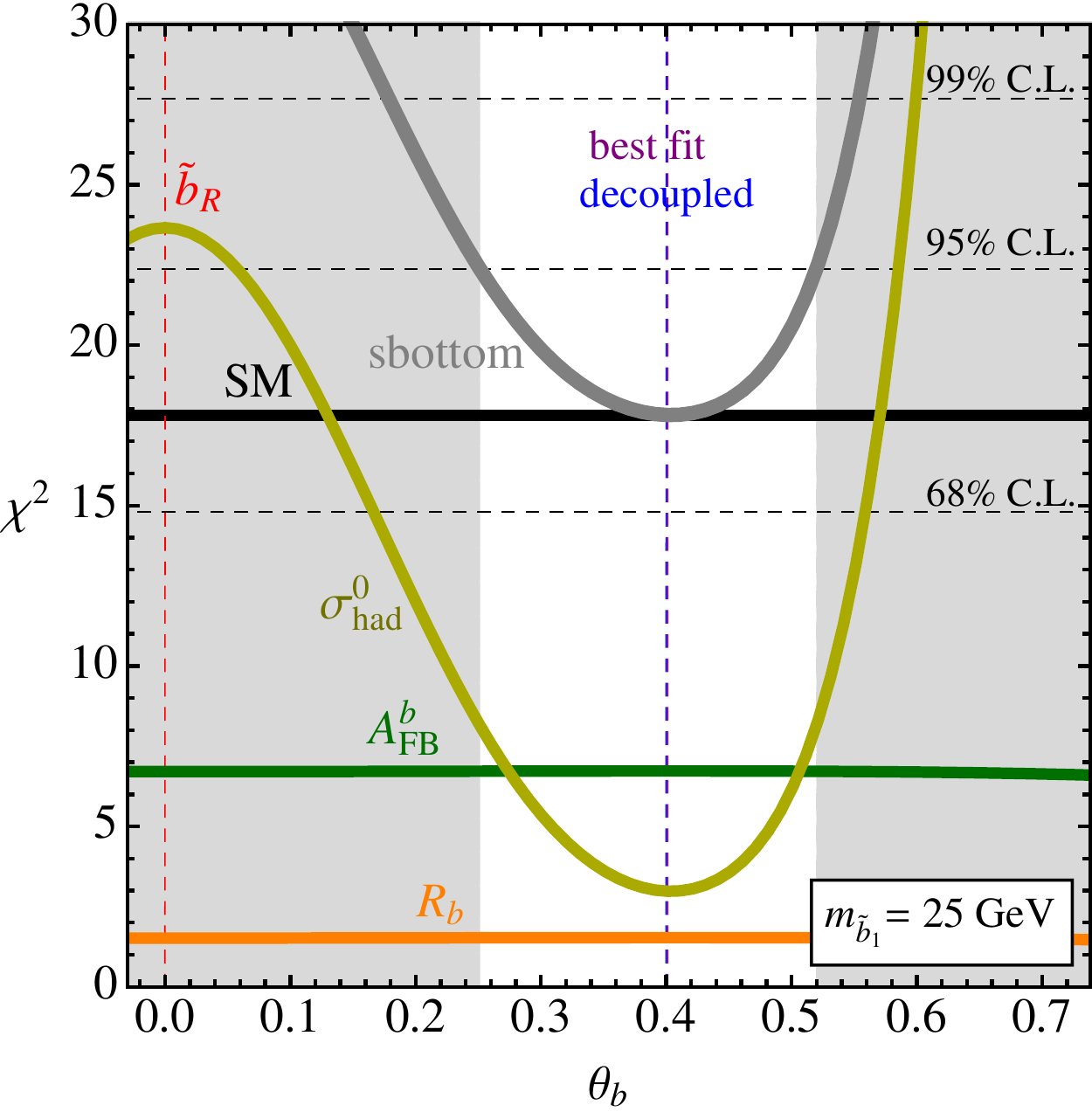} ~~~
\includegraphics[width=0.45\textwidth]{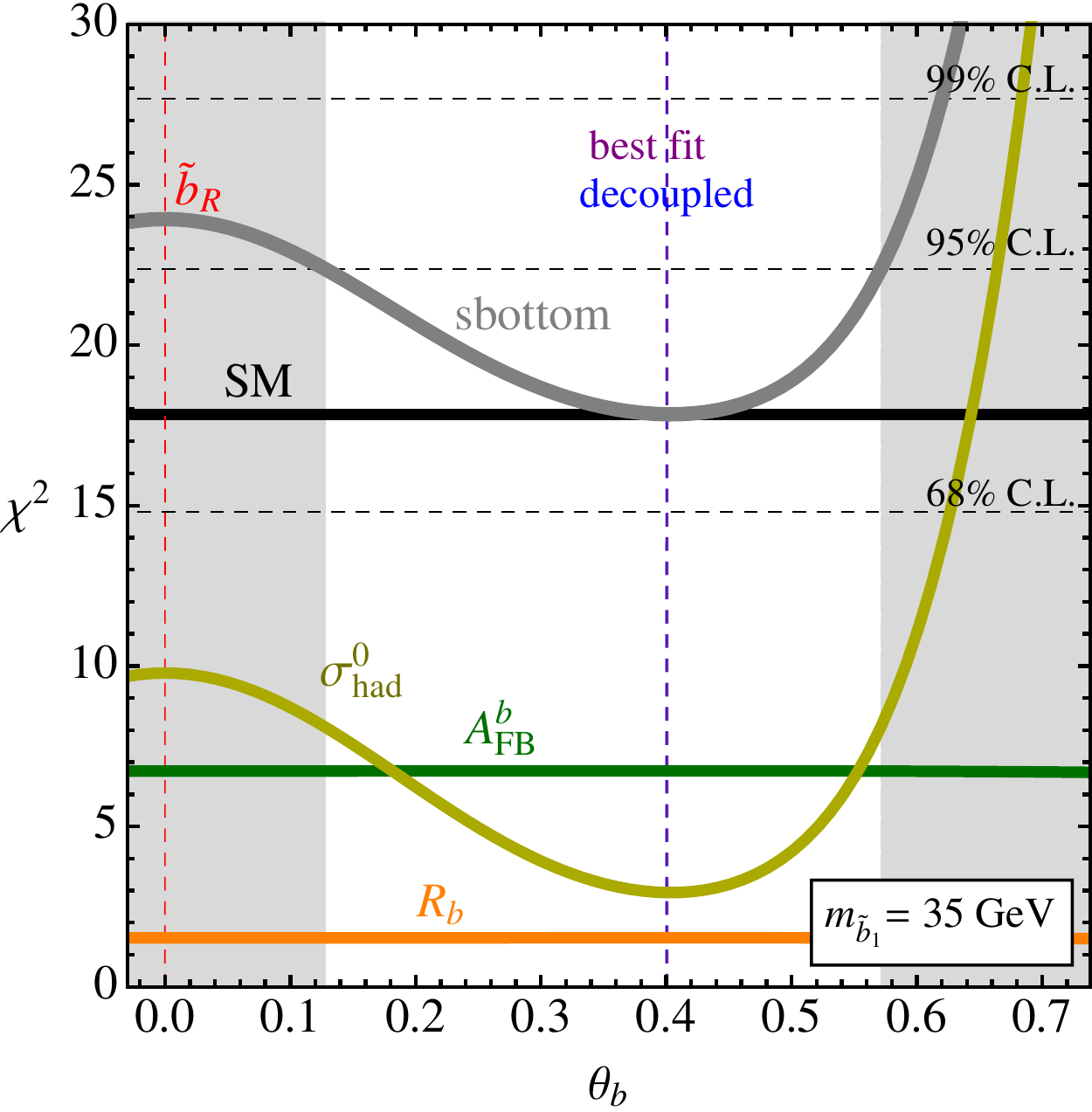} 
\caption{ \small
{\it Global fit to precision electroweak data}. Here we display the results of the fit for the case that the sbottom is ``invisible''. We show the $\chi^2$ statistic for the global fit to a light sbottom (solid grey) and the contribution to $\chi^2$ from the observables $\sigma_{\rm had}^0$ (solid yellow), $R_b$ (solid orange)  and $A_{FB}^b$ (solid green) 
as a function of the mixing angle $\theta_b$. 
The results are displayed for three values of the sbottom mass: $m_{\tilde b_1} = 5.5, 15, 25, 35$ GeV.
We also display the $68,95,99\%$ C.L. values (dashed black) for $\nu = 13$ degrees of freedom (19 observables, 6 fit parameters), as well as the SM value $\chi^2_{\rm SM} = 17.8$  (solid black) for comparison. The grey shaded regions display a tension with the precision electroweak data, with a $p$ value less than $0.05$. Finally, we represent the special cases of a pure right-handed sbottom $\tilde b_R$  ($\theta_b = 0$, dashed red), a decoupled sbottom ($\theta_b = 0.4$, dashed blue), and the best fit values (dashed purple).}
\label{fig:ewpd-inv}
\end{center}
\end{figure*}

\begin{figure*}
\begin{center}
\includegraphics[width=0.45\textwidth]{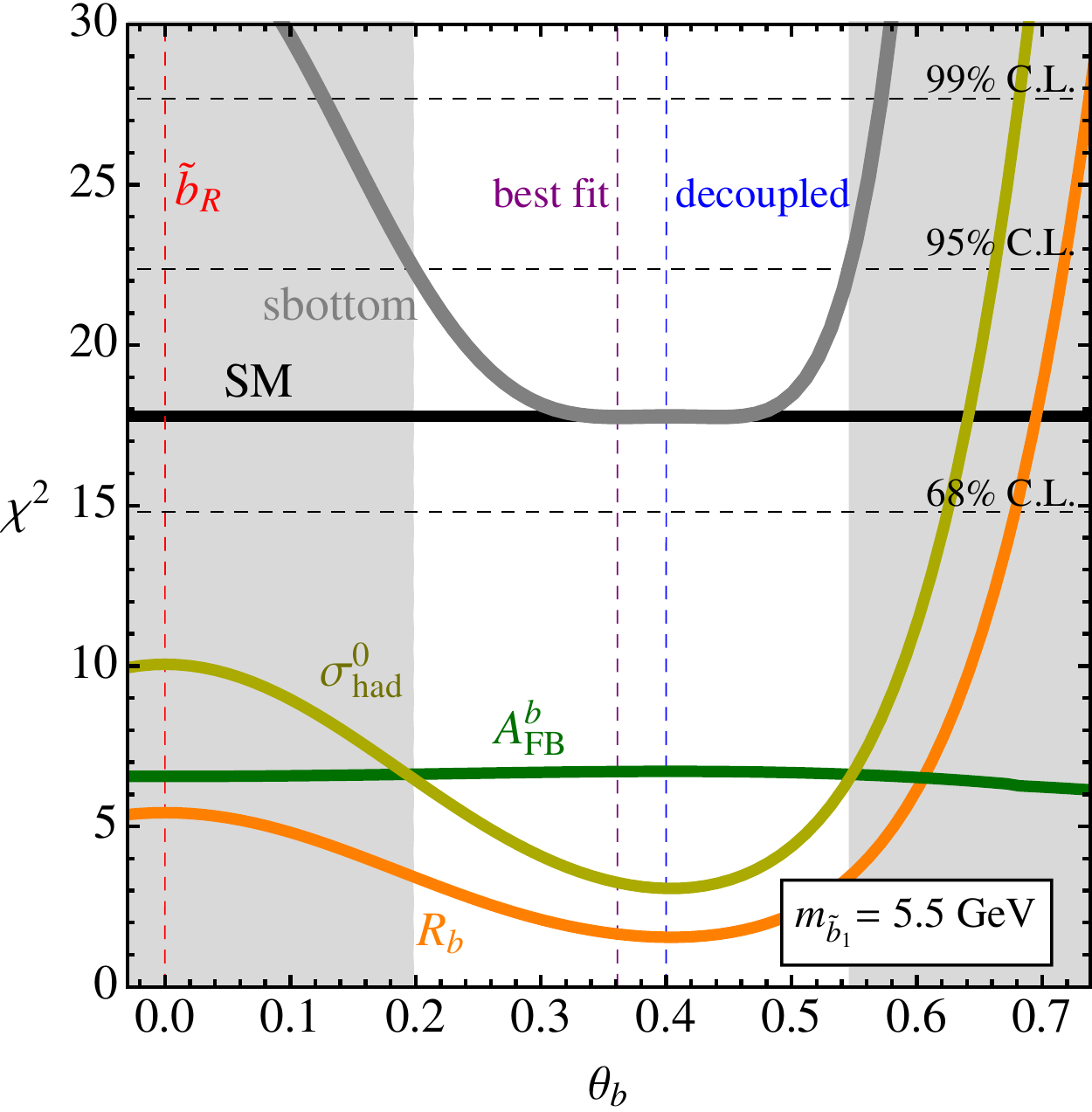}~~~
\includegraphics[width=0.45\textwidth]{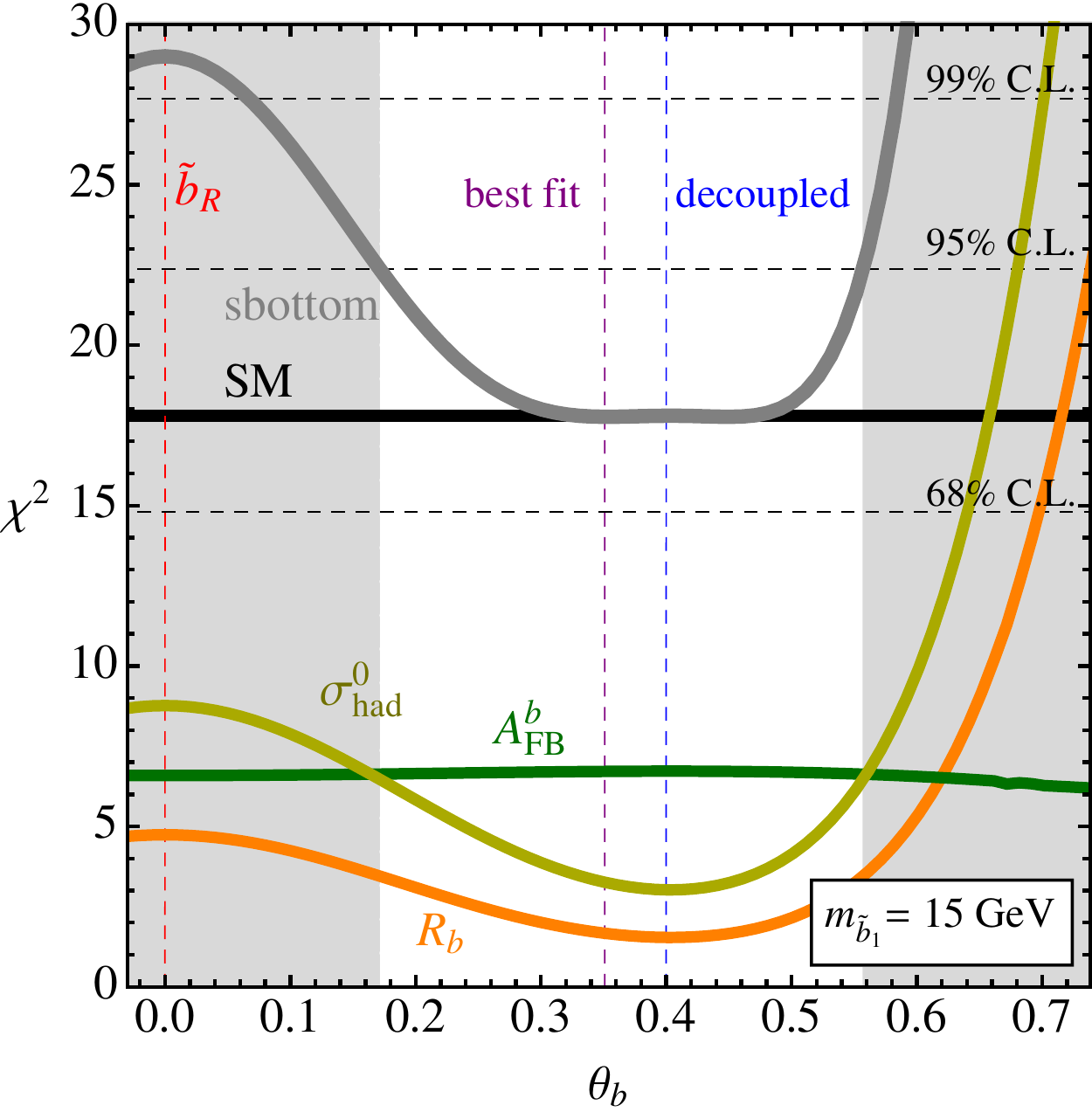} \\
\includegraphics[width=0.45\textwidth]{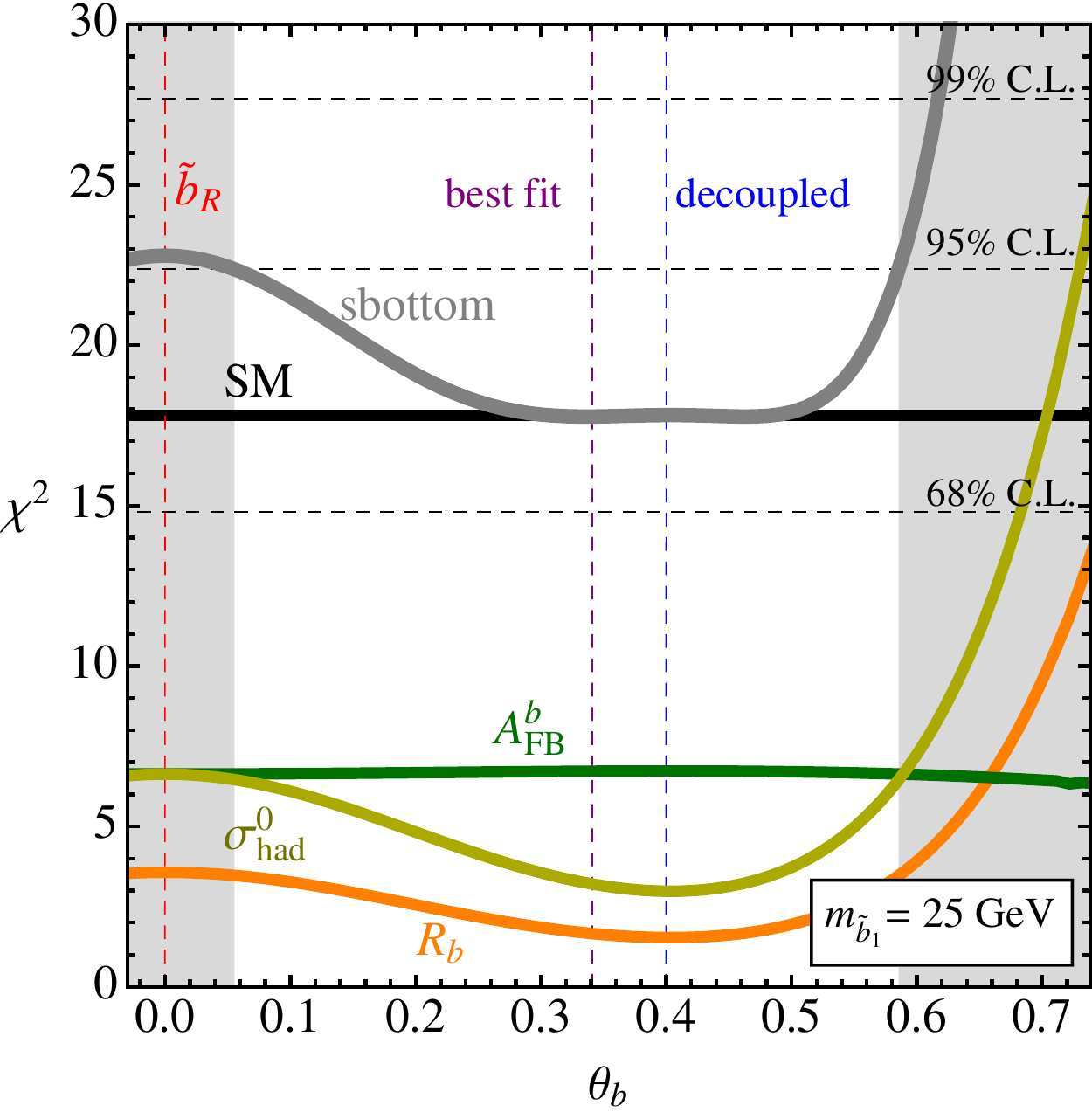} ~~~
\includegraphics[width=0.45\textwidth]{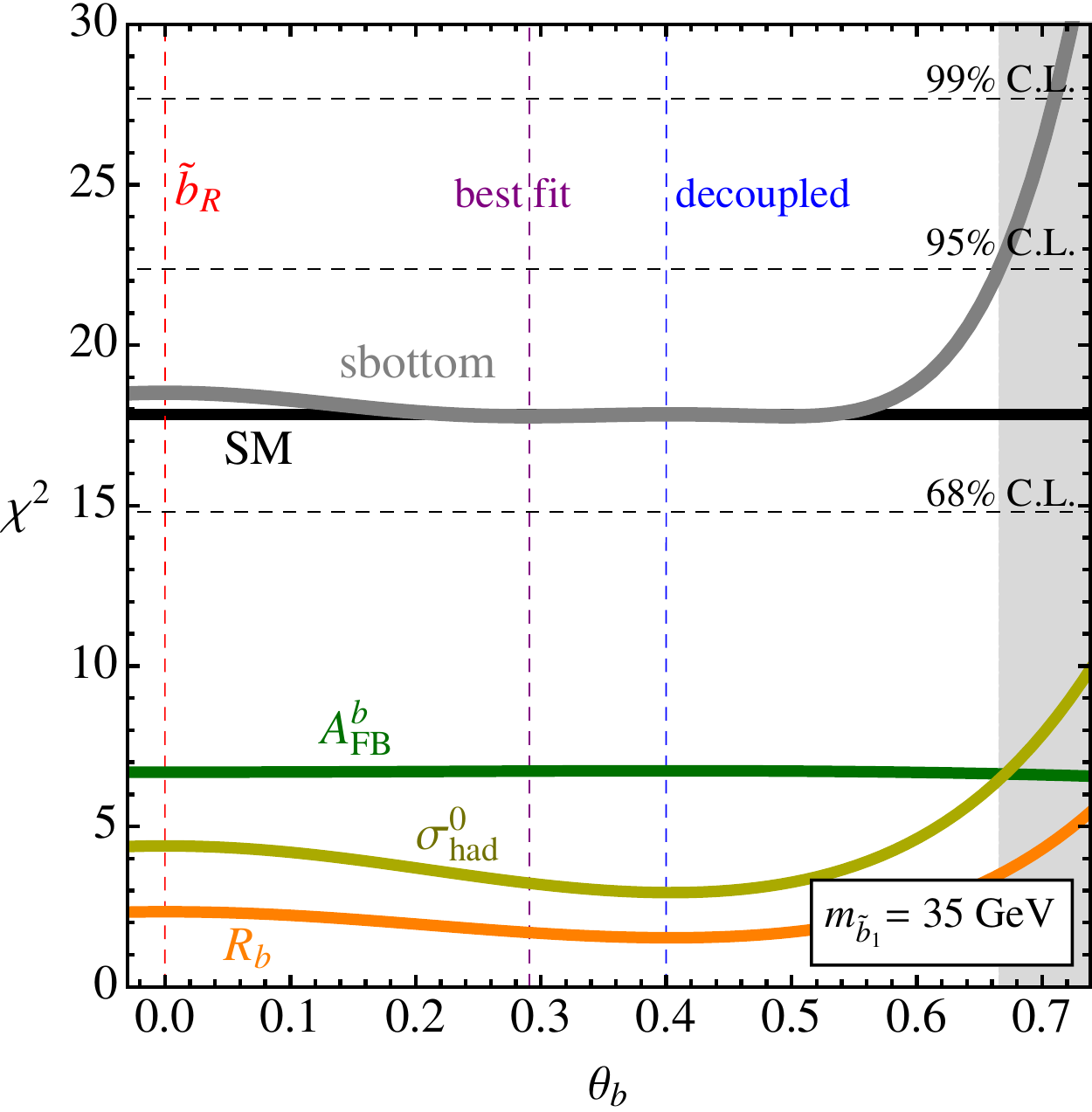} 
\caption{ \small
As in Fig.~\ref{fig:ewpd-inv}, but for the case that the sbottom is reconstructed as hadrons.}
\label{fig:ewpd-had}
\end{center}
\end{figure*}

\begin{figure*}
\begin{center}
\includegraphics[width=0.45\textwidth]{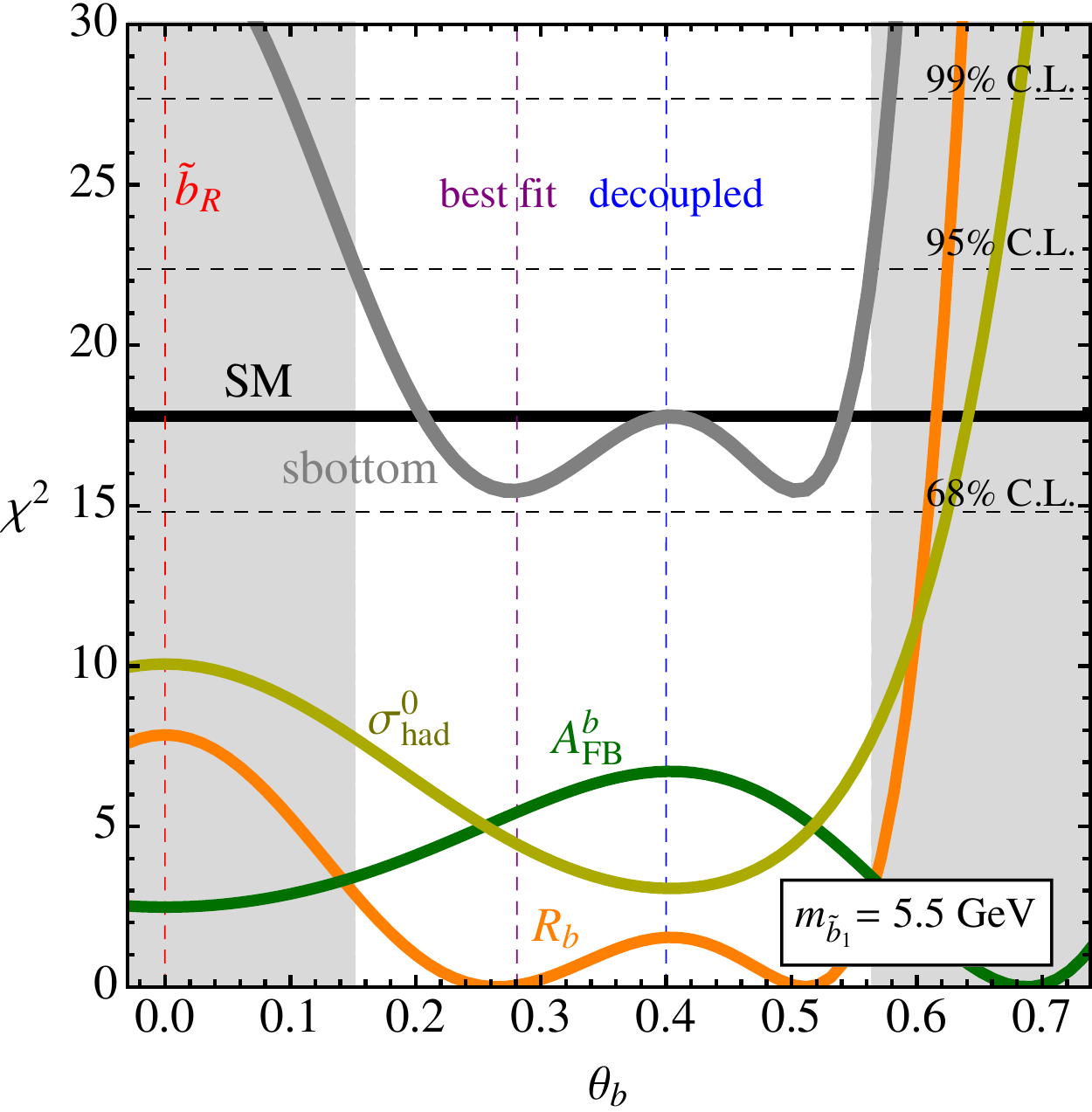}~~~
\includegraphics[width=0.45\textwidth]{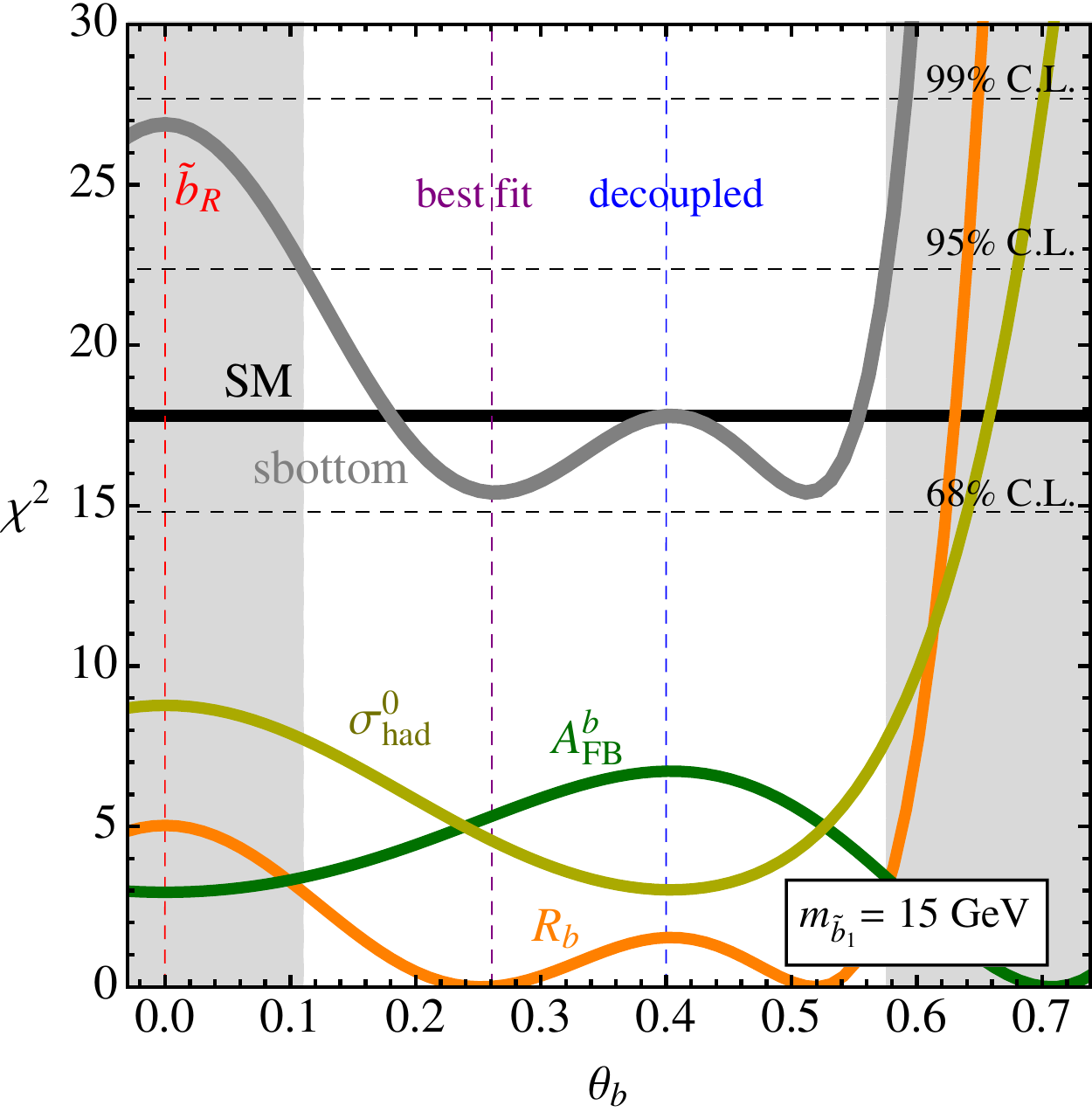} \\
\includegraphics[width=0.45\textwidth]{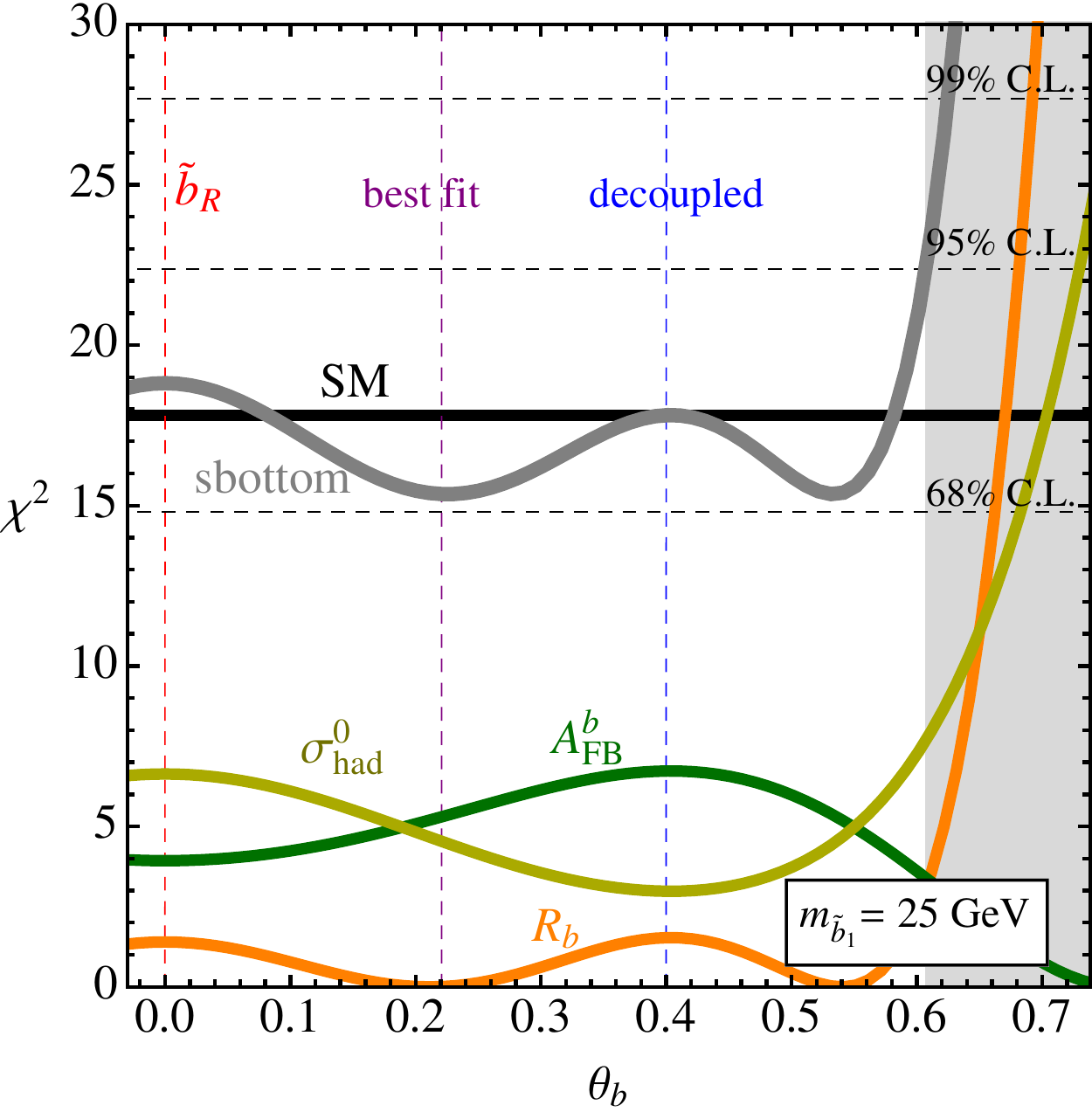} ~~~
\includegraphics[width=0.45\textwidth]{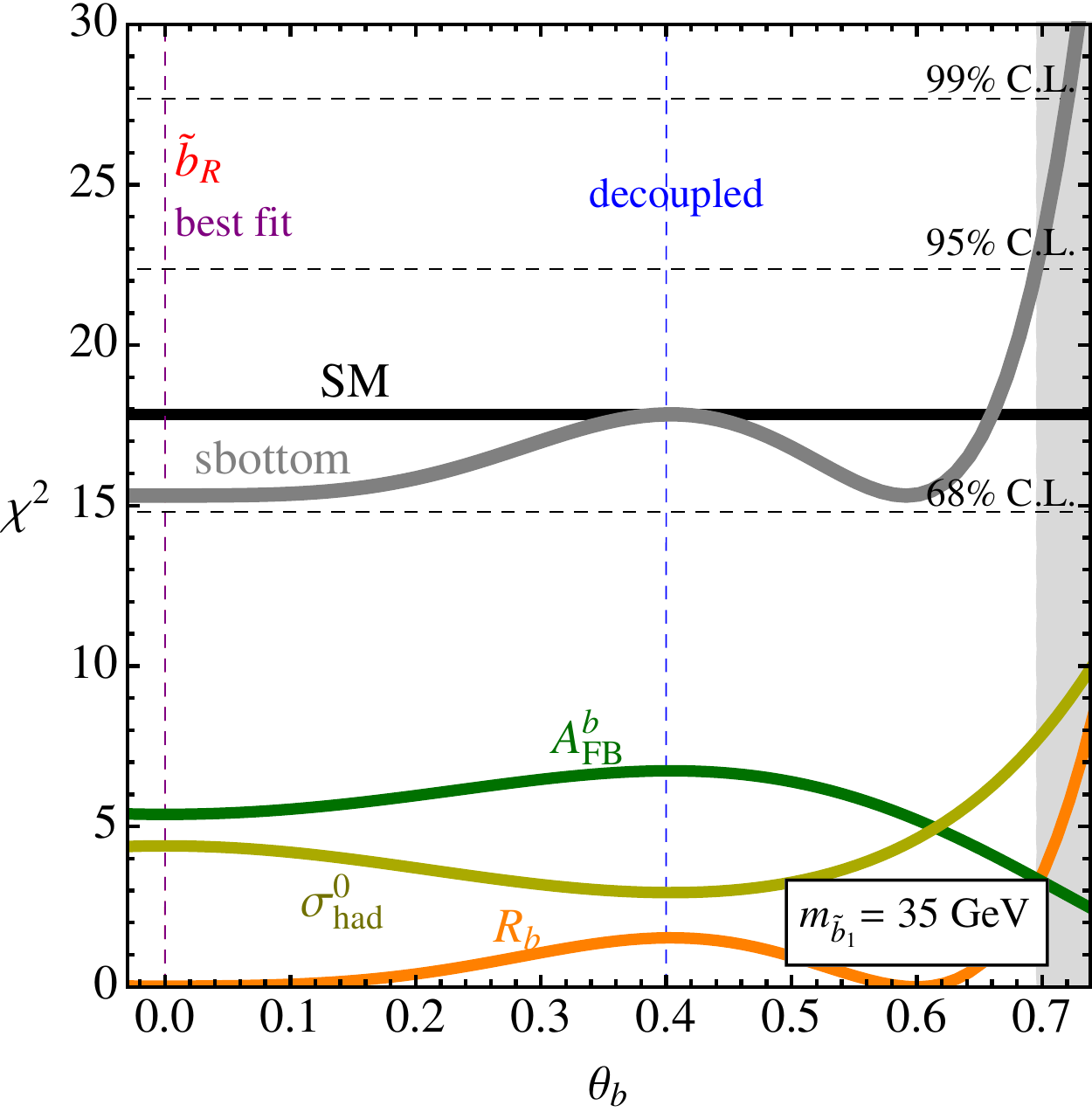} 
\caption{ \small
As in Fig.~\ref{fig:ewpd-inv}, but for the case that the sbottom is reconstructed as a $b$ quark.}
\label{fig:ewpd-b}
\end{center}
\end{figure*}

In Figs.~\ref{fig:ewpd-inv},\ref{fig:ewpd-had},\ref{fig:ewpd-b} we present the results of the global electroweak fit to the light sbottom scenario for each of the three sbottom reconstruction hypothesis presented at the beginning of Sec.~\ref{sec:EWPD}. Results are presented for four different sbottom masses, $m_{\tilde b_1} =$ (5.5, 15, 25, 35) GeV.
These plots display the global $\chi^2$ statistic (solid grey) as a function of the sbottom mixing angle. For each sbottom mixing angle we marginalize over the SM fit parameters. Additionally, in each plot the $68,95,99\%$ C.L. values (dashed black) for $\nu = 13$ degrees of freedom (19 observables, 6 fit parameters), as well as the SM value $\chi^2_{\rm SM} = 17.8$  (solid black) are displayed for comparison. The shaded grey regions exhibit a tension with the data at the $95\%$ C.L. while the white regions provide an acceptable description of the data. We refer the reader to Secs.~\ref{subsec:sbottom-EWPD} for the predictions to the various observables. 

First, as a reference point we give the results for the fit to the SM. The SM fit (with $\alpha_s$ determined by the fit) has 18 observables and 5 fit parameters for 13 d.o.f., and yields $\chi^2_{\rm SM} = 17.8$, which corresponds to a $p$-value of 0.17, in good agreement with the results of Gfitter~\cite{Baak:2012kk}. The SM therefore provides an acceptable description of the precision data. The SM value $\chi^2_{\rm SM}$ is displayed in Figs.~\ref{fig:ewpd-inv},\ref{fig:ewpd-had},\ref{fig:ewpd-b} by the solid black line.

Next we examine the fit results for a light sbottom assuming that it is ``invisible", {\it i.e.}, not counted in visible channels, which are displayed in Fig.~\ref{fig:ewpd-inv}.
Especially for very light sbottoms, we observe that consistency with the data requires a mixing angle that is close to the decoupling value, $\theta_b \sim 0.4$. Away from this value, {\it e.g.}, for mostly right-handed sbottoms ($\theta_b \sim 0$) or for highly mixed and left-handed sbottoms, the total peak hadronic cross section $\sigma_{\rm had}^0$ is much too small compared to the measured value, yielding a poor global fit. Another general feature that is obseved in Fig.~\ref{fig:ewpd-inv} is that the allowed parameter space opens up for larger sbottom masses, which can easily be understood as a consequence of the phase space suppression in the $Z\rightarrow \tilde b_1 \tilde b_1^*$ width in Eq.~(\ref{eq:Z-Sbottom-decay}).

For the case of the hadronic sbottom, the results of the fit are presented in Fig.~\ref{fig:ewpd-had}. As in the case of the ``invisible'' sbottom just discussed, there is a strong preference for a mixing angle near the decoupling value $\theta_b \sim 0.4$. Large mixing angles $\theta_b \gtrsim 0.7$ are disfavored regardless of the sbottom mass, while for light sbottoms below about 25 GeV, purely right-handed sbottoms $\theta_b \sim 0$ are also disfavored. Again, the main culprit is $\sigma_{\rm had}^0$, which becomes much smaller than the experimental value for mixing angles away from the decoupling value. Note that in comparision to the ``invisible" sbottom there is a broader region around 
the decoupling value that provides a good description of the data. This is because the increase in $\Gamma_{\rm had}$ as $\theta_b$ becomes smaller raises the prediction for $R_\ell$, bringing it in better agreement with the measured value. This partly compensates the increased tension in $\sigma_{\rm had}^0$ and $R_b$ as $\theta_b$ is decreased.

Lastly, we consider the case in which the sbottom is counted as a $b$ quark. The results for this case are presented in Fig.~\ref{fig:ewpd-b}. Again, for very light sbottoms below about 15 GeV, there is  a window around the decoupling value $\theta_b \sim 0.4$ that provides a good description of the data. 
Interestingly, if the sbottom is counted as a $b$ quark, the global fit can actually be improved with respect to the SM, as the predictions for $R_b$ and $A_{FB}^b$ are in better agreement with the measured values for mixing angles $\theta_b \sim 0.2-0.3$ and $\theta_b \sim 0.5$. This is indicated by the two local minima in the total $\chi^2$ (solid grey) in
Fig.~\ref{fig:ewpd-b}. We also observe that the tension in $A_{FB}^b$ can be moderately reduced ({\it e.g.}, the disagreement is at the 1.8$\sigma$ level for $m_{\tilde b_1} = 5.5$ GeV, $\theta_b  = 0.15$) although the discrepancy cannot be fully explained by this hypothesis without spoiling the agreement  in $\sigma_{\rm had}^0$. For heavier sbottoms, about about 25 GeV, the region near $\theta_b \sim 0$ also is allowed. We stress, however, that it is likely challenging to construct a scenario in which such a heavy $\tilde b_1$ can mimic a $b$-quark, since the emitted $b$ quarks may be softer or more acoplanar than if  directly produced. On the other hand, it is possible for a light $\tilde b_1$ to mimic a $b$ quark, and we will discuss such a scenario in Sec.~\ref{sec:collider}.

%%%%%%%%%

%%%%%%%%%

%%%%%%%%%

\subsection{Summary}

We have seen that the precision electroweak data impose strong constraints on the possible existence of a very light sbottom. Regardless of how the sbottom is counted, light sbottoms with mass below 15 GeV must have a mixing angle near the decoupling value $\theta_b \sim 0.4$ in order to provide an acceptable global description of the data. In particular, purely right-handed sbottoms $\theta_b \sim 0$ and  highly mixed or left handed sbottoms $\theta_b \gtrsim 0.7$ are generically in tension with the precision data for 
$m_{\tilde b_1} \lesssim 15$ GeV.  Purely right handed sbottoms can be consistent for larger masses due the phase space suppression in the  $Z \rightarrow \tilde b_1 \tilde b_1^*$ width. 
The main observable responsible for the constraints is the total hadronic cross section at the $Z$-peak, $\sigma_{\rm had}^0$. 
This SM prediction for $\sigma_{\rm had}^0$ is smaller than the experimental value by about 1.5$\sigma$, and a light sbottom can only lower the prediction and thus increase the tension.

 A summary of our results are presented in Table~\ref{tab:EWPD-summary}, where we display the $95\%$ C.L. allowed range for the mixing angle $\theta_b$, the $p$-value for the best-fit mixing angle, and for comparison the $p$-value for a purely right-handed sbottom $\theta_b=0$. The results are displayed for $m_{\tilde b_1} =$ 5.5, 15, 25, 35 GeV for each sbottom reconstruction hypothesis. 
 
%%%%%%%%%
\begin{table}
\begin{center}
\renewcommand{\arraystretch}{0.7}
 \begin{tabular}{| c || c | c | c | | c | c | c || c | c | c ||}
\hline
  & \multicolumn{3}{|c||}{~1. ``Invisible"}
 & \multicolumn{3}{|c||}{~2. Hadronic} 
 & \multicolumn{3}{|c||}{~3. $b$ ~~~}  
 \\  \hline
%%%%%%%%%%%%%%
%%%%%%%%%%%%%%
\multirow{2}{*}{ \, $m_{\tilde b_1}$    \, } 
&  $\theta_b$  
& Best fit 
& RH
&  $\theta_b$  
& Best fit 
& RH
&  $\theta_b$  
& Best fit 
& RH
\\ 
%\cline{2-3} 
&  ($95\,\%$C.L.)   
& $p$-value  
&$p$-value  
&  ($95\,\%$C.L.)   
& $p$-value  
&$p$-value  
&  ($95\,\%$C.L.)   
& $p$-value  
&$p$-value  
 \\ \hline
%%%%%%%%%%%%%%
5.5 GeV & [0.29,0.49]   &  0.17 & $\sim10^{-8}$  & [0.20,0.55]   &  0.17 & 0.001  & [0.15,0.56]   &  0.28 & 0.002  \\ \hline
%%%%%%%%%%%%%%
15 GeV & [0.28,0.50]   &  0.17 & $\sim10^{-6}$  & [0.17,0.56]   &  0.17 & 0.007  & [0.11,0.58]   &  0.28 & 0.013  \\ \hline
%%%%%%%%%%%%%%
25 GeV & [0.25,0.52]   &  0.16 &  $0.0004$  & [0.06,0.59]   &  0.17 & 0.044  & [0,0.61]   &  0.29 & 0.13  \\ \hline
%%%%%%%%%%%%%%
35 GeV & [0.13,0.57]   &  0.16 &  $0.03$  & [0,0.67]   &  0.17 & 0.14  & [0,0.70]   &  0.29 & 0.13  \\ \hline
\hline
  \end{tabular}
\end{center}
\caption{
{\it Summary of the global fit to the precision electroweak data. }
We display the $95\%$ C.L. allowed range for the mixing angle $\theta_b$, the $p$-value for the best-fit mixing angle, and for comparison the $p$-value for a purely 
right-handed sbottom $\theta_b=0$. 
The results are displayed for $m_{\tilde b_1} =$ 5.5, 15, 25, 35 GeV for each sbottom reconstruction hypothesis, which are described in detail at the beginning of Sec.\ref{sec:EWPD}. 
}
\label{tab:EWPD-summary}
\end{table}

\newpage

%%%%%%%%%%%%%%%%%%%%%%%%%%%%%%
%%%%%%%%%%%%%%%%%%%%%%%%%%%%%%
\section{Higgs Signal Strength Data}\label{sec:higgs}

The 126 GeV Higgs $h^0$ can decay to light sbottom pairs, $h^0 \rightarrow \tilde b_1 \tilde b_1^*$ and, if kinematically allowed, $h^0 \rightarrow \tilde b_1 \tilde b_2^*$. 
There are already non-trivial bounds on the branching ratios to these modes arising from the signal strength measurements of the Higgs boson. 
The main effect of these new decay modes is to dilute the branching fractions of the most precisely measured channels,  $\gamma\gamma$, $ZZ$, and $WW$,  which are broadly in accord with the SM predictions. Beyond this effect, one must again understand how the sbottom is counted in Higgs searches. For instance, if the sbottom mimics the $b$ quark through its decay, then there will also be an apparent increase in the branching ratio in the $b \bar b$ channel.

To quantify the impact of the sbottoms on the Higgs signal strength data, we perform a fit using the results of Ref.~\cite{Belanger:2013xza}. In that work, from the raw ATLAS, CMS, and Tevatron data the authors derive a $\chi^2$ function for eight channels, one for each combination of two combined production modes and four final states. The combined production modes are 1) gluon-gluon fusion and $t-t-h$ (ggF+ttH), and 2) vector boson fusion and associated production (VBF+VH), while the final states considered are $\gamma\gamma$, $VV$, $b\bar b$, and $\tau \bar \tau$. 

With the current level of precision in the $b \bar b$ signal strength measurement, we expect that the constraints will not be very sensitive to how the sbottom is counted. 
To confirm this, we have investigated in a model independent fashion the allowed size of a new contribution to the $h\rightarrow b\bar b$ partial width 
$\delta \Gamma(h\rightarrow b\bar b)$ as well as the invisible width  $\Gamma(h\rightarrow {\rm invisible})$, using two fits to the Higgs signal strength data. 
In the first fit, we add a new contribution to the $h\rightarrow b\bar b$ partial width and determine $\delta \Gamma(h\rightarrow b\bar b) < 1.6$ MeV at $95\%$ C.L. In the second fit, we consider a new invisible width, obtaining $\Gamma(h\rightarrow {\rm invisible}) < 1.7$ MeV at $95\%$ C.L. The fact that these limits are so similar confirms our expectation that the principal effect of the new decay modes is to dilute the $\gamma\gamma$, $ZZ$, and $WW$ signal strengths. 
For simplicity, in the numerical results below we have assumed that the new decay modes of the Higgs to sbottoms are invisible. 

While the most important constraint comes from the the new decay mode $h^0 \rightarrow \tilde b_1 \tilde b_1^*$, another effect that  can be numerically important 
is the modification of the the $h\rightarrow gg$ partial width (and thus the gluon fusion cross section) and to a lesser degree the $h\rightarrow \gamma\gamma$ partial width from new one loop diagrams involving sbottom and stop exchange. 

In the next two subsections we will describe in detail the new contributions to the Higgs decay modes and the $h\rightarrow gg$, $h\rightarrow \gamma\gamma$ partial  widths.

\subsection{New Higgs decay modes}\label{subsec:higgsdecay}
Since the mass of the lightest sbottom under consideration is well below $m_{h^0}/2$, there is a new decay mode $h^0 \rightarrow \tilde b_1 \tilde b_1^*$. The partial width for this decay is given by 
\begin{equation}
\Gamma(h^0 \rightarrow \tilde b_1 \tilde b_1^*)  = \lambda_{h^0 \tilde b_1 \tilde b_1^*}^2  \frac{3}{16 \pi m_{h^0}} \left( 1 - \frac{ 4 m_{\tilde b_1}^2  }{m_{h^0}^2}  \right)^{1/2}.
\label{eq:higgs-sbottoms}
\end{equation}
In the decoupling limit ($m_{A^0} \gg m_Z$), the coupling of the Higgs boson to the lightest sbottom mass eigenstate is given by (see also the Appendix 
(\ref{eq:sfermion-higgs}))
\begin{eqnarray}
\label{eq:higgs-sbottom-coupling}
 \lambda_{h^0 \tilde b_1 \tilde b_1^*}  & = &
\sqrt{2} v \bigg\{      
\frac{m_b^2}{v^2} +
 \frac{m_Z^2 \cos{2 \beta} }{v^2} \big[  s^2_{b}\left(- \tfrac{1}{2} + \tfrac{1}{3} s_W^2  \right)  +     c^2_{b}\left(- \tfrac{1}{3} s_W^2  \right)  \big]
   + c_{b} s_{b} \frac{m_b (A_b - \mu \tan\beta)}{v^2} 
\bigg\},~~~~~~~
\end{eqnarray} 
where $v = 174$ GeV, $s_b \equiv \sin\theta_b$, etc. 

\begin{figure*}
\begin{center}
\includegraphics[width=0.6\textwidth]{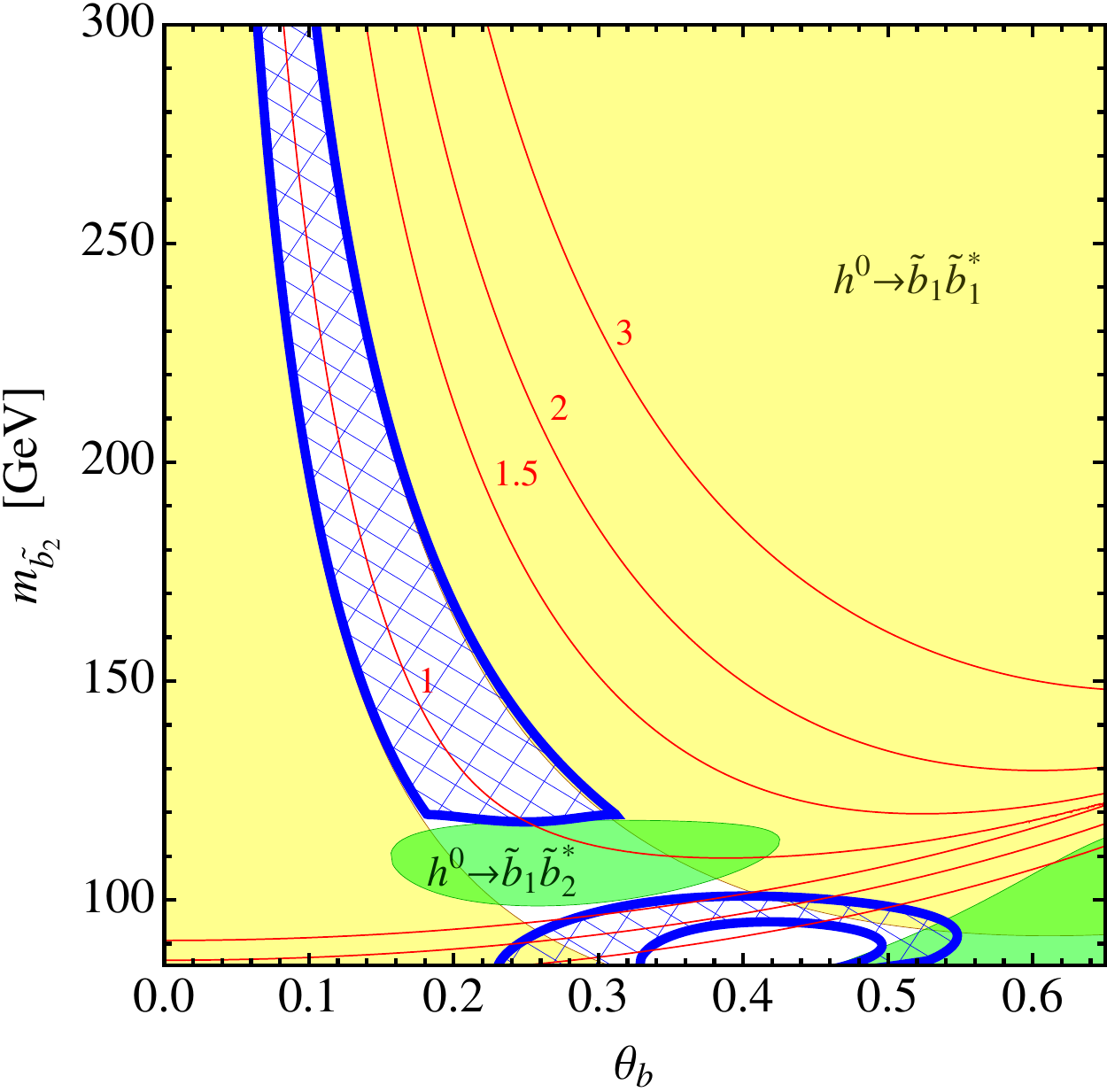}~~~
\caption{ 
\small  
{\it Fit to the Higgs signal strength data}:  
The blue hatched area indicates the region of $\theta_b - m_{\tilde b_2}$ parameter space that provides an acceptable 
fit to the Higgs signal strength data set.  
The yellow (green) shaded region is excluded by considering {\it only} the new decay $h^0 \rightarrow \tilde b_1 \tilde b_1^*$ ($h^0 \rightarrow \tilde b_1 \tilde b_2^*$).
The red contours represent the enhancement in the $h\rightarrow gg$ partial width compared to the SM prediction. The correction to the $h\rightarrow \gamma\gamma$ rate is small, of order 5-10$\%$ over the allowed parameter space, and is not indicated in the plot. See the text for a detailed explanation of the shape of the allowed region (blue hatched).  In this example we have fixed the following inputs:
$\mu = 200$ GeV, 
$\tan\beta = 10$, 
$m_{\tilde b_1} = 5.5$ GeV, 
$m_{U_3} = 2.5$ TeV, 
$A_t = 2$ TeV. 
}
\label{fig:Higgs-SS}
\end{center}
\end{figure*}

This new decay mode is already strongly constrained by the Higgs signal strength data. In Fig.~\ref{fig:Higgs-SS} we display the allowed parameter space in the 
$\theta_b - m_{\tilde b_2}$  plane, for the inputs 
$\mu = 200$ GeV, 
$\tan\beta = 10$, 
$m_{\tilde b_1} = 5.5$ GeV, 
$m_{U_3} = 2.5$ TeV, 
$A_t = 2$ TeV. 
The region in yellow is excluded by considering {\it only} the effect of the decay $h^0 \rightarrow \tilde b_1 \tilde b_1^*$ on the signal strength data. 
The thin white strip corresponds to the region where the coupling $\lambda_{h^0 \tilde b_1 \tilde b_1^*}$ in Eq.~(\ref{eq:higgs-sbottom-coupling}) is small and the decay is suppressed. 
Using the tree level relation between Lagrangian and physical parameters in the sbottom sector,
\begin{equation}
\label{eq:tree-relation}
m_b(A_b - \mu \tan\beta) = s_b\, c_b\, (m_{\tilde b_1}^2 - m_{\tilde b_2}^2),
\end{equation} 
we observe that the coupling approximately vanishes along the curve
\begin{eqnarray}
\label{eq:tune-mb-thetab}
m_{\tilde b_2} & \simeq & 
m_Z \sqrt{\cos {2\beta}  \big[  \sec^2{\theta_b} \left(- \tfrac{1}{2} + \tfrac{1}{3} s_W^2  \right)  +   \csc^2{\theta_b} \left(- \tfrac{1}{3} s_W^2  \right)  \big]}  \\
&  \approx   &  \frac{m_Z s_W}{\sqrt{3}\, \theta_b}, ~~~~(\theta_b \ll 1, \tan\beta \gg 1).  \nonumber 
\end{eqnarray}
This relation (\ref{eq:tune-mb-thetab})
must be obeyed to a good approximation in order to evade the constraints from the Higgs data. 

If the second sbottom is light enough, $m_{\tilde b_2} < m_{h^0}-m_{\tilde b_1}$, then the Higgs can also decay via $h^0 \rightarrow \tilde b_1 \tilde b_2^*, \tilde b_1^* \tilde b_2$.
The partial decay widths are 
\begin{equation}
\Gamma(h^0 \rightarrow \tilde b_1 \tilde b_2^*) = \Gamma(h^0 \rightarrow \tilde b_1^* \tilde b_2)  = 
 \lambda_{h^0 \tilde b_1 \tilde b_2^*}^2  \frac{3}{16 \pi m_{h^0}}  \lambda^{1/2}\left(1,  \frac{m_{\tilde b_1}^2  }{m_{h^0}^2} , \frac{m_{\tilde b_2}^2  }{m_{h^0}^2}  \right),
\label{eq:higgs-sbottoms}
\end{equation}
where $\lambda(a,b,c) = a^2 + b^2 + c^2 - 2 a b -2 a c -2 b c$, and the coupling $\lambda_{h^0 \tilde b_1 \tilde b_2^*}$ is given by (in the decoupling limit)
\begin{eqnarray}
\label{eq:higgs-sbottom-coupling-2}
\lambda_{h^0 \tilde b_1 \tilde b_2^*} & = & 
\frac{v}{\sqrt{2}}  \left\{  \sin{2\theta_b} \frac{m_Z^2 \cos{2\beta}}{v^2} \left( -\tfrac{1}{2}+\tfrac{2}{3}s_W^2  \right)
+ \cos{2\theta_b} \frac{m_b (A_b - \mu \tan\beta)}{v^2}    \right\}.
\end{eqnarray} 
where $v = 174$ GeV.  
The green regions in Fig.~(\ref{fig:Higgs-SS}) are excluded by considering {\it only} the effect of the decays 
$h^0 \rightarrow \tilde b_1 \tilde b_2^*,  \tilde b_1^* \tilde b_2$ on the signal strength predictions. 
The first thing to observe is that the constraints vanish for second sbottom masses heavier than about 120 GeV, corresponding to the value of $m_h^0 - m_{\tilde b_1}$ in this example.
One can understand the allowed regions by again focusing on where the coupling (\ref{eq:higgs-sbottom-coupling-2}) is small. Using the relation (\ref{eq:tree-relation}), we find that the coupling $\lambda_{h^0 \tilde b_1 \tilde b_2^*}$  (\ref{eq:higgs-sbottom-coupling-2}) is proportional to $\sin{2\theta_b}$ and thus vanishes as $\theta_b \rightarrow 0$, explaining why there are no constraints in this region. Furthermore, there is a second region where the coupling is small, centered around the curve
\begin{equation}
m_{\tilde b_2}  = m_Z \sqrt{ \frac{\cos2\beta(-1+\tfrac{4}{3} s_W^2)}{\cos{2\theta_b}}}.
\end{equation}
which is situated between the two green regions in Fig.~(\ref{fig:Higgs-SS}). 

As one can observe from Fig.~(\ref{fig:Higgs-SS}), the new contributions to the gluon fusion cross section can shift the allowed parameter space as the enhancement in this cross section can compensate to a certain degree the dilution in the branching ratios to the $\gamma\gamma, WW, ZZ$ channels. We discuss these effects next.  

\subsection{Modifications to $h^0\rightarrow gg, \gamma\gamma$}
While the new decay modes of the Higgs to light sbottoms are the dominant factor governing the constraints on the sbottom parameter space, the modifications to the loop induced Higgs couplings, particularly to the gluons, do make a quantitative impact. This is because an enhancement in the gluon fusion rate can offset the effect of a new decay channel and restore the signal strength predictions to SM-like values. 

The complete formulae for the one loop corrections to the $h\rightarrow gg, \gamma\gamma$ partial widths are presented in the Appendix~\ref{appendix} and are used to obtain our numerical results. In 
Fig.~(\ref{fig:Higgs-SS}) we have displayed red isocontours of the enhancement factor $r_{g}  = \Gamma(h^0 \rightarrow gg)/ \Gamma(h^0 \rightarrow gg)_{\rm SM}$. For the parameters chosen in this plot, the stop contribution to $r_g$ is negligible. In fact, as we will discuss in the next section, this is motivated by requirement of a small $\Delta \rho$ parameter, which typically coincides with a small coupling of the Higgs to the lightest stop. 

The dominant modifications to the gluon fusion rate originate from the diagrams containing sbottoms. As is clear from Fig.~(\ref{fig:Higgs-SS}) there are two distinct behaviors of the contours depending on whether $m_{\tilde b_2}$ is larger or smaller than $m_{h^0}$. In the regime  $m_{\tilde b_2} \gtrsim m_{h^0}$, the lightest sbottom $\tilde b_1$ gives the largest correction, which can be written as
\begin{eqnarray}
r_{g} &\simeq & 
\Bigg\vert  \,
1 \, +
 \, \frac{1}{\sqrt{2} A_{gg}^{\rm SM}} \,
 \frac{v \, \lambda_{h^0 \tilde b_1 \tilde b^*_1} }{ m_{\tilde b_1}^2 }   
 A_0(m_{\tilde b_1})   
\Bigg\vert^2,   \nonumber \\
& \approx & \Bigg\vert  \,
1 \, +
 \, \frac{\sin^2{2\theta_b}}{ A_{gg}^{\rm SM}} \,
 \frac{m_{\tilde b_2}^2 }{ m_{h^0}^2 },   
\Bigg\vert^2.
\end{eqnarray}
In the second line above we have kept only the leading terms from the general formulae for $r_g$ (\ref{eq:rg}), the sbottom-Higgs coupling $ \lambda_{h^0 \tilde b_1 \tilde b^*_1}$ (\ref{eq:sfermion-higgs}), and the scalar loop function, 
\begin{equation}
A_0(m_{\tilde b_1}) \simeq - \frac{4m_{\tilde b_1}^2}{ m_{h^0}^2}, ~~~~~ (m_{\tilde b_1} \ll m_{h^0}   )
\end{equation}
 and employed the relation 
(\ref{eq:tree-relation}) to write the last formula above in terms of the physical sbottom mass $m_{\tilde b_2}$ and mixing angle $\theta_b$. This approximation is valid for moderate mixing angles, {\it i.e.}, to the right of the white band in Fig.~(\ref{fig:Higgs-SS}). We can see that for a fixed value 
$r_g$ traces out the curve $m_{\tilde b_2} \sim \sqrt{A_{gg}^{\rm SM} (\sqrt{r_g}-1) } m_{h^0}/2\theta_b$. 
Again, this approximation is valid for $m_{\tilde b_2} \gtrsim m_{h^0}$.

Finally, there are also modifications to the $h\rightarrow \gamma\gamma$ rate, although in comparison with $h\rightarrow gg$, the effects are very small. Over the range of viable parameter space where the new decays of the Higgs to sbottoms are suppresed, we find that there is a small $5-10\%$ suppression in the partial width. 

\subsection{Summary}

Putting together the effects of the new Higgs decay modes and the one loop modifications to the $h\rightarrow gg, \gamma\gamma$ couplings, we obtain in 
Fig.~(\ref{fig:Higgs-SS}) the blue hatched region, which represents the parameters allowed by the Higgs signal strength data set at $95\%$ C.L. 
(or equivalently, the parameters that yield a $p$-value greater than 0.05 in the global fit). 
When the second sbottom is heavy, $m_{\tilde b_2} \gtrsim m_{h^0}$, the allowed region is governed entirely by the requirement that the decay $h^0 \rightarrow \tilde b_1 \tilde b_1^*$ is suppressed, as discussed in Sec.~\ref{subsec:higgsdecay}.
Near masses $m_{\tilde b_2} \sim 100$ GeV, the presence of the new decay modes $h^0 \rightarrow \tilde b_1 \tilde b_2^*,b_1^* \tilde b_2$ causes a dilution of the
branching ratios of the Higgs decays to SM particles, although the corresponding production modes are compensated to a certain degree by an enhancement in the gluon fusion cross section\footnote{We note that if the second sbottom is very light, $m_{\tilde b_2}  < m_{Z} - m_{\tilde b_1}$, there will be further precision electroweak constraints coming from the new decay modes of the $Z$ boson, $Z \rightarrow \tilde b_1 \tilde b_2^*, \tilde b_1^* \tilde b_2$.}. There is also a ``hole'' that is excluded near $m_{\tilde b_2} \sim 90$ GeV, $\theta_b \sim 0.4$ because the gluon fusion rate is too large, and the new decays of the Higgs to sbottoms are suppressed. 

In the example above, we have chosen parameters in the stop sector such that the coupling of the lightest stops to the Higgs is small. This is motivated by the phenomenological constraint of the $\Delta \rho$ parameter, which affects the precision observables. In the next section, we well explore this requirement in more detail, examining how a 
a non-zero $\Delta \rho$ parameter affects the fits to the precision electroweak data.

\section{Effects of the stops}\label{sec:stops}
In the previous two sections we have examined in detail the effects of light sbottoms on the precision electroweak and Higgs signal strength datasets.  
As alluded to in the previous section, it is also important to investigate the effects of the stops for several reasons. 
First, depending on parameters in the stop sector, there can be a large custodial symmetry breaking and thus  a $\Delta \rho$ parameter
which will alter the predictions for the precision observables. 
Furthermore, if we restrict to the MSSM, then in order to obtain the observed value of the Higgs mass, the second stop should be fairly heavy, and there should be a tight correlation between the soft parameters $m_{U_3}$ and $A_t$. Finally, the stops can also contribute to the one loop decays 
$h\rightarrow gg, \gamma\gamma$, as discussed in the previous section. 

\subsection{$\Delta \rho$}

Here we consider the implications of a nonzero contribution to $\Delta \rho$ to the precision electroweak observables. The predictions for these observables in the presence of 
$\Delta T \equiv \Delta \rho/\alpha$ can be obtained from Ref.~\cite{Burgess:1993vc}, which we have incorporated into our global fit. We also include the contributions of the light sbottom, as detailed in Sec.~\ref{sec:EWPD}.

\begin{figure*}
\begin{center}
\includegraphics[width=0.45\textwidth]{chi2sbottom-b-6GeV.pdf}~~~
\includegraphics[width=0.45\textwidth]{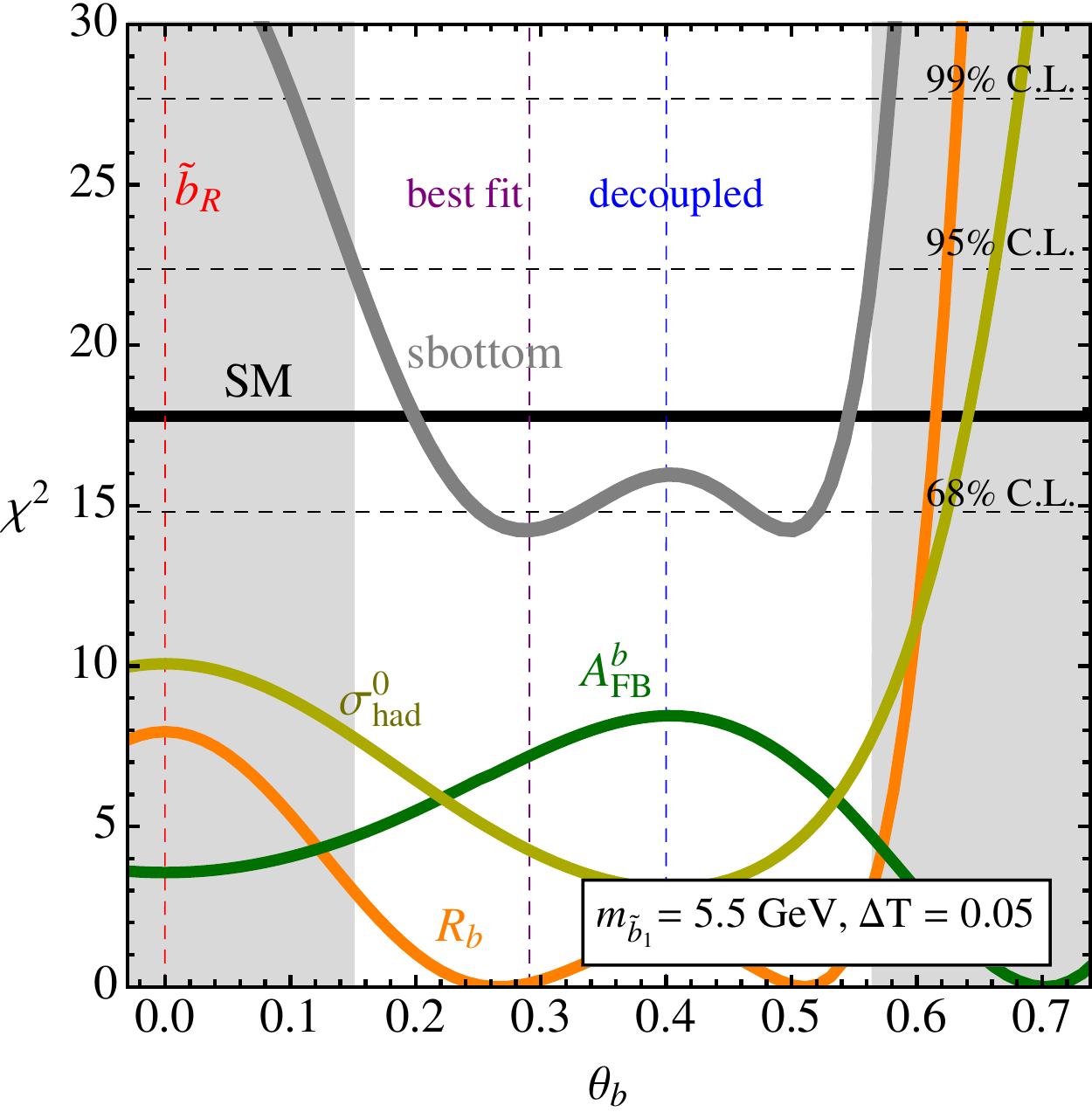} \\
\includegraphics[width=0.45\textwidth]{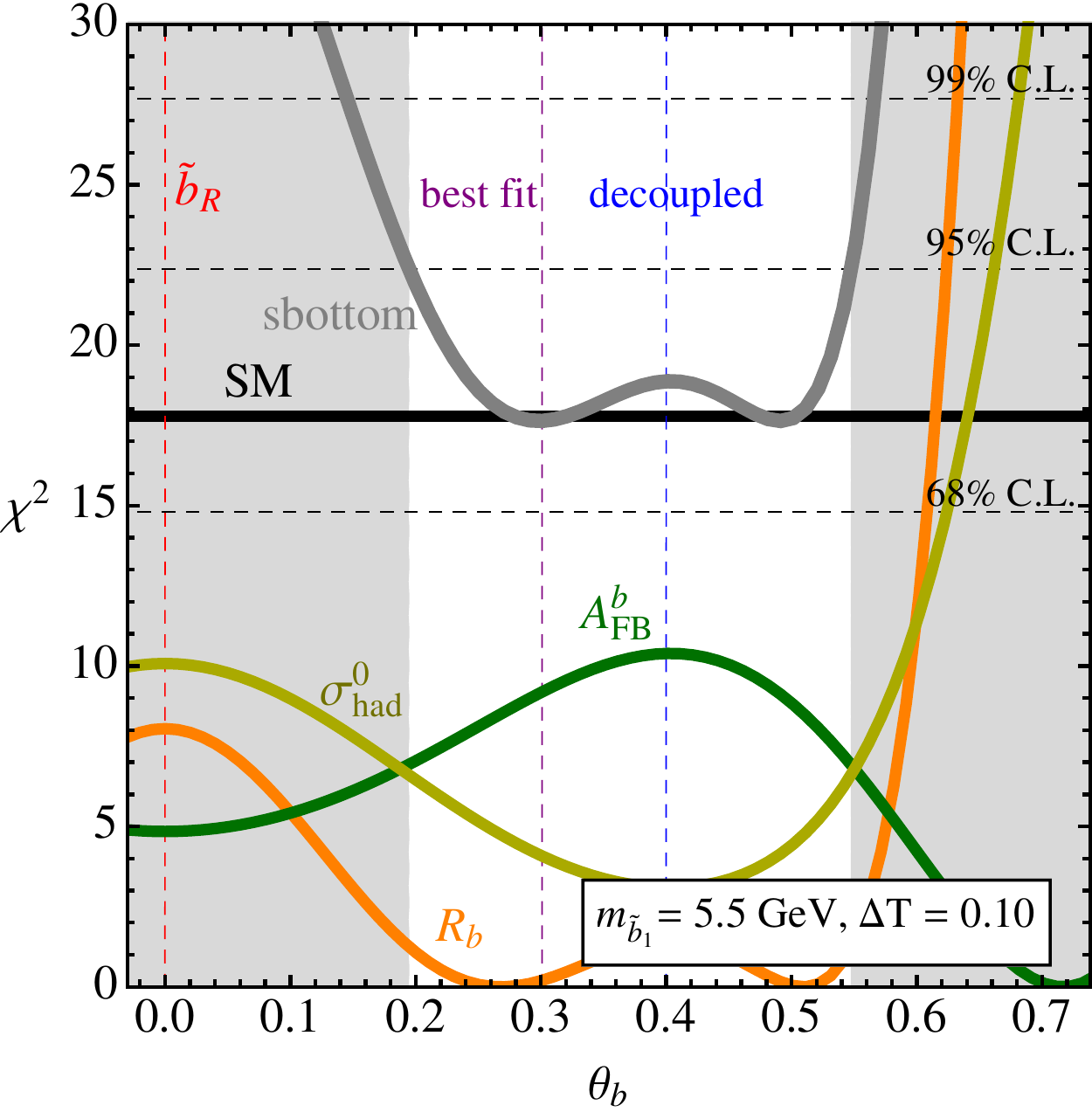} ~~~
\includegraphics[width=0.45\textwidth]{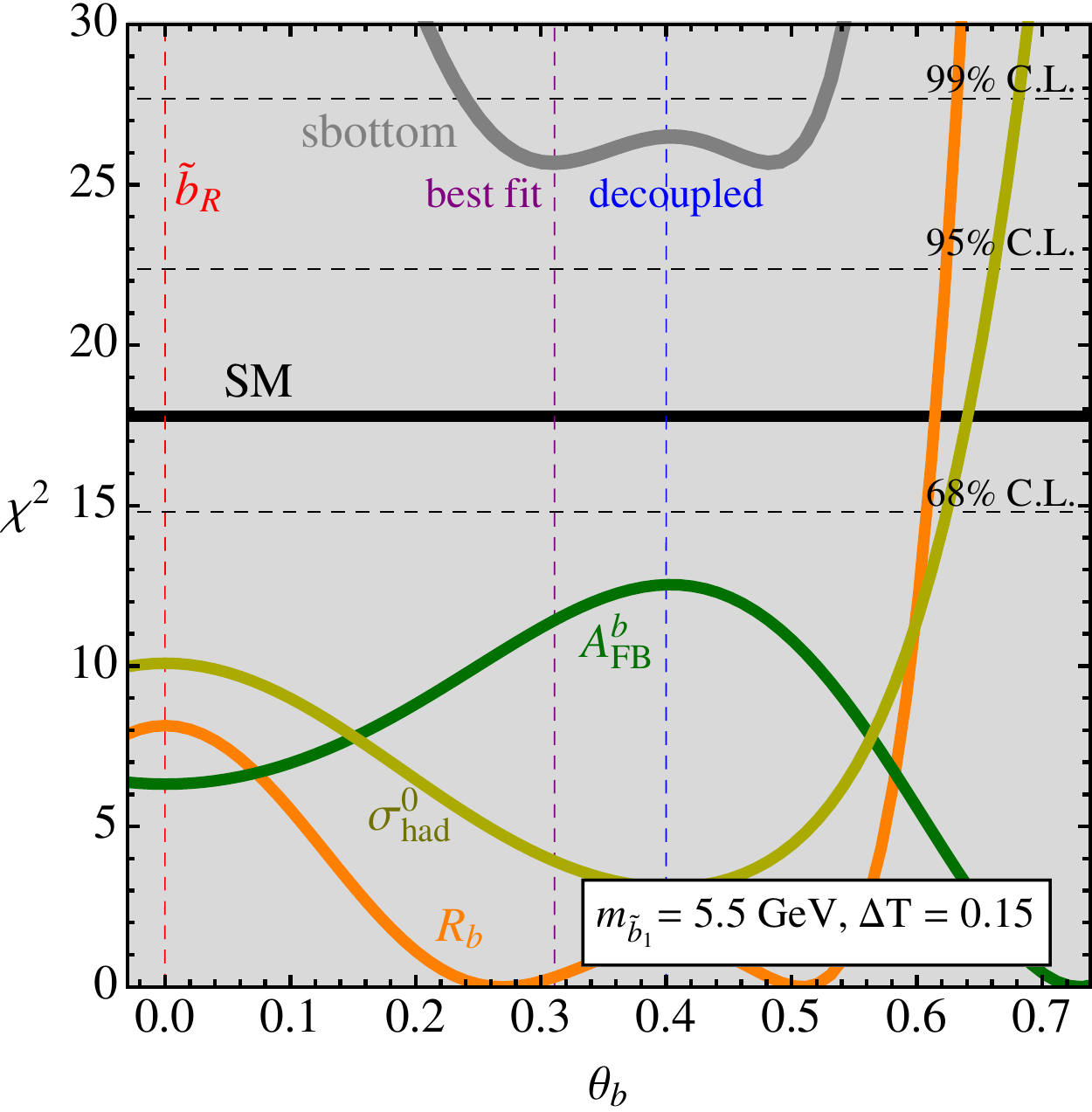} 
\caption{ \small
{\it Effects of $\Delta \rho$ on the precision electroweak data}: 
Here we show the $\chi^2$ statistic for the global fit to a 5.5 GeV sbottom (solid grey), here assumed to be counted as a $b$ quark,  and the contribution to $\chi^2$ from the observables $\sigma_{\rm had}^0$ (solid yellow), $R_b$ (solid orange)  and $A_{FB}^b$ (solid green) 
as a function of the mixing angle $\theta_b$. 
The results are displayed for four values of $\Delta T = \Delta \rho /\alpha$:~~ $\Delta T = 0, 0.05, 0.1, 0.15$.
We also display the $68,95,99\%$ C.L. values (dashed black) for $\nu = 13$ degrees of freedom (19 observables, 6 fit parameters), as well as the SM value $\chi^2_{\rm SM} = 17.8$  (solid black) for comparison. The grey shaded regions display a tension with the precision electroweak data, with a $p$ value less than $0.05$. Finally, we represent the special cases of a pure right-handed sbottom $\tilde b_R$  ($g_{Z \tilde b_1 \tilde b_1^*} = s_W^2/3$, dashed red), a decoupled sbottom ($g_{Z \tilde b_1 \tilde b_1^*} = 0$, dashed blue), and the best fit values (dashed purple).}
\label{fig:ewpd-T}
\end{center}
\end{figure*}

In Figure.~\ref{fig:ewpd-T} we illustrate the effects of a non-zero $\Delta \rho$ parameter on the fit. Here we have taken a lightest sbottom mass of $m_{\tilde b_1} = 5.5$ GeV, and assumed it is counted as a $b$-quark. This is the most optimistic scenario, while the cases of ``invisible'' and hadronic sbottoms are more constrained; see 
Sec.~\ref{sec:EWPD} for a discussion. In these fits, we have treated $\Delta T$ as a free parameter and have shown the results for four values $\Delta T = 0, 0.05, 0.1, 0.15$. A small $\Delta \rho$ parameter gives a comparable fit to the case of $\Delta \rho = 0$. For example, in the case of $\Delta T = 0.05$ the fit is slightly improved as the $W$ boson mass prediction is in better agreement with the measured value. However, one clearly observes that as $\Delta T$ becomes larger than about 0.1, there is increasing tension in the fit, 
which is driven mainly by the observables $A_{FB}^b$, $m_W$, $\Gamma_Z$, $\sigma_{\rm had}^0$. 

\begin{figure*}
\begin{center}
\includegraphics[width=0.6\textwidth]{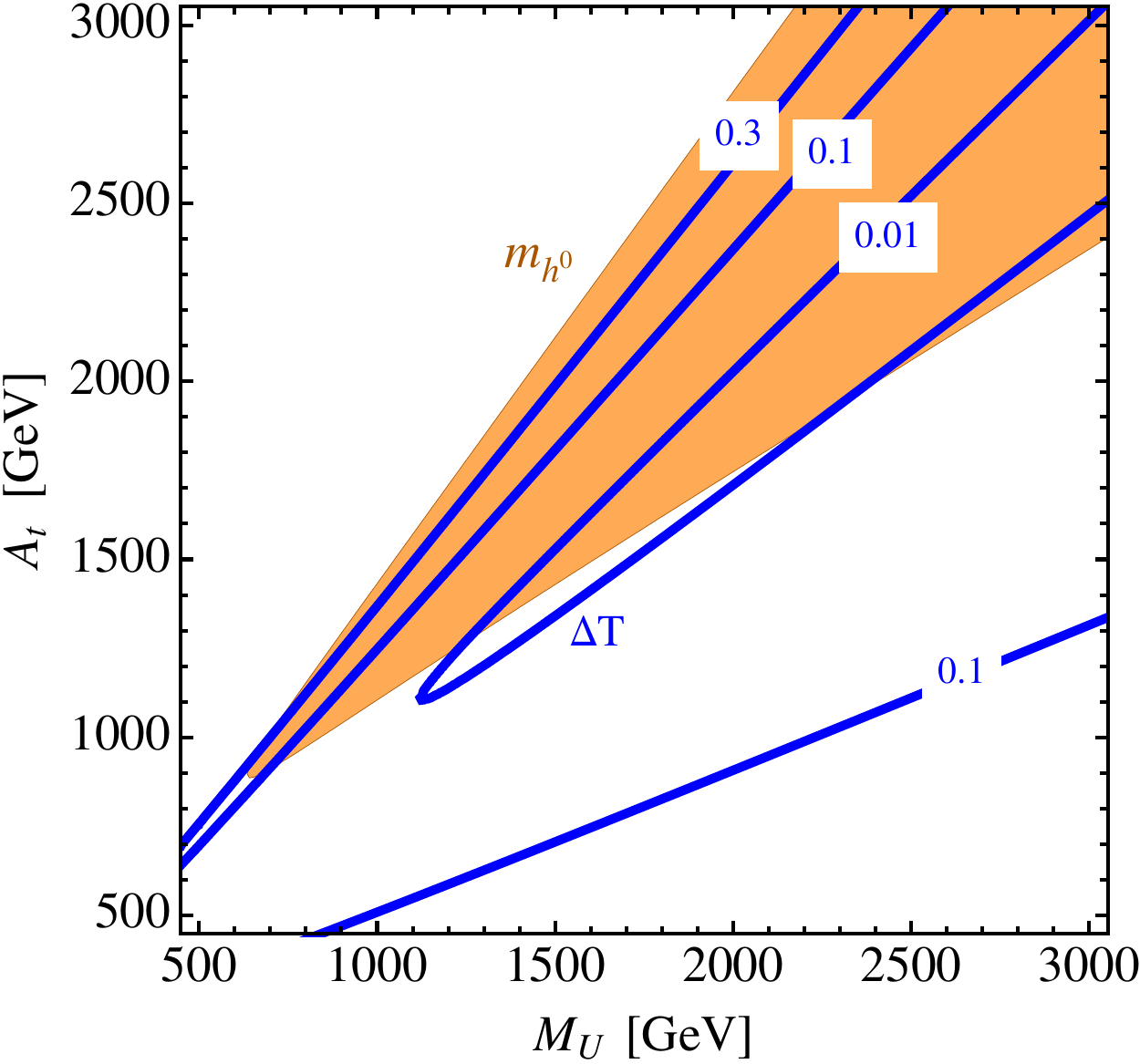}~~~
\caption{ 
\small  
{\it Higgs mass and $\Delta \rho$ parameter}:  Here we display in the $m_{U_3} - A_t$ plane the region where the Higgs mass 
predicted by Eq.~(\ref{eq:mh2loop-Ap}) lies in the range $122\,{\rm GeV} \leq m_{h^0} \leq 128\,{\rm GeV}$ (orange). We have also overlaid isocontours of $\Delta T = \Delta \rho /\alpha$ (blue). In this example we have taken the inputs 
$\mu = 200$ GeV, 
$\tan\beta = 10$, 
$m_{\tilde b_1} = 5.5$ GeV, 
$m_{\tilde b_2} = 200$ GeV, 
and
$\theta_b = 0.15$.
}
\label{fig:HiggsMass-rho}
\end{center}
\end{figure*}

Next, we turn to the contributions of the stop - sbottom sector. The one loop contribution to $\Delta \rho$ can be found in Ref.~\cite{Heinemeyer:2004gx}, and is given by
\begin{eqnarray}
\Delta \rho & = &  \frac{3 G_F}{8 \sqrt{2}\, \pi^2}   
\bigg[
- c_t^2 s_t^2 F(m_{\tilde t_1}^2, m_{\tilde t_2}^2) 
- c_b^2 s_b^2 F(m_{\tilde b_1}^2, m_{\tilde b_2}^2) 
+ s_t^2 s_b^2 F(m_{\tilde t_1}^2, m_{\tilde b_1}^2)  
\nonumber \\ 
&&
~~~~~~~~~~\,\,
+ s_t^2 c_b^2 F(m_{\tilde t_1}^2, m_{\tilde b_2}^2)  
+ c_t^2 s_b^2 F(m_{\tilde t_2}^2, m_{\tilde b_1}^2)  
+ c_t^2 c_b^2 F(m_{\tilde t_2}^2, m_{\tilde b_2}^2)  
 \bigg], 
\end{eqnarray}
where the function $F$ is defined as
\begin{equation}
F(x,y) = x+y-\frac{2\,x\,y}{x-y}\log{\frac{x}{y}}.
\end{equation}
In Fig.~\ref{fig:HiggsMass-rho} we display in the $m_{U_3} - A_t$ plane isocontrours of $\Delta T$, for the inputs $\mu = 200$,  $\tan\beta = 10$, $m_{\tilde b_1} = 5.5$ GeV,  $m_{\tilde b_2} = 200$ GeV, $\theta_b = 0.15$.  We observe that the $T$ parameter is minimized as the second stop mass, $m_{\tilde t_2} \sim m_{U_3}$ becomes larger, 
and for values of $A_t \sim m_{U_3}$. The reason for the latter is related to custodial symmetry breaking: In the limit, $m_{U_3}, A_t \gg m_{Q_3}, m_t \gg m_Z$, the lightest stop is primarily left-handed, with squared mass given by $m_{\tilde t_1}^2 \simeq  m_{Q_3}^2 + m_t^2 (1-A_t^2/m_{U_3}^2)$, while  the second sbottom is mostly left-handed with mass $m_{\tilde b_2} \simeq m_{Q_3}$. Therefore, the custodial symmetry breaking, and hence $\Delta \rho$ is minimized for parameters $A_t \approx m_{U_3}$.

\subsection{Higgs mass}

In the MSSM, large contributions from the stops are required to raise the Higgs mass from the tree level value, which is less than $m_Z$, to the observed value of 126 GeV. To estimate the range of parameter space that gives the correct Higgs mass, we employ the following two loop approximate formula appropriate for a hierarchical stop spectrum:
\begin{eqnarray}
\label{eq:mh2loop-Ap}
m_{h^0}^2   &   \approx  & m_Z^2 \cos^2{2\beta} + \frac{3}{8 \pi^2} \frac{m_t^4}{v^2}  
\Bigg[ \tilde X_t + t_1+t_2 \nonumber  
  +  \frac{1}{16 \pi^2} \bigg\{  \tilde X_t \bigg[  \frac{m_t^2}{v^2} \left( 8 t_2 -5 t_1 \right) - 64 \pi \alpha_3 t_2   \bigg] \\ &&
 + \, \frac{m_t^2}{v^2} \left[ 4(t_1^2 + t_2^2) -5 t_1 t_2 \right]  - 32 \pi \alpha_3 (t_1^2 + t_2^2)  
  \bigg\}
 \Bigg], \nonumber 
\end{eqnarray}
where  $m_t$ is the running top quark mass, and we have defined
\begin{equation}
t_{1,2} \equiv \log{\frac{m_{\tilde t_{1,2}}^2}{m_t^2}}, ~~~\tilde X_t \equiv \frac{2 X_t^2}{m_{\tilde t_2}^2 -m_{\tilde t_1}^2} \log{\frac{m_{\tilde t_{2}}^2}{m_{\tilde t_{1}}^2}} 
+  \frac{X_t^4}{(m_{\tilde t_2}^2 -m_{\tilde t_1}^2)^2} \left[ 2 - \frac{ m_{\tilde t_{2}}^2 + m_{\tilde t_{1}}^2  }{ m_{\tilde t_{2}}^2 - m_{\tilde t_{1}}^2  }
\log{\frac{m_{\tilde t_{2}}^2}{m_{\tilde t_{1}}^2}} 
  \right].
 \end{equation}
This equation contains the dominant leading log effects from the stops. We are neglecting subdominant effects from two-loop thresholds, sbottoms loops, and electroweak corrections. We have performed numerical checks with FeynHiggs~\cite{Frank:2006yh} and CPsuperH~\cite{Frank:2006yh,Lee:2003nta} and find agreement to within about 3 GeV.
Note that in the limit $m_{\tilde t_1} \rightarrow m_{\tilde t_2}$ reproduces the analogous approximate formula for the case of degenerate stops from Ref.~\cite{Carena:1995wu}. 

In Fig.~\ref{fig:HiggsMass-rho} we represent in orange the region of parameter space that yields $122$ GeV $< m_h^0 < 128$ GeV (accounting for the $O({\rm few ~ GeV})$  uncertainty in Eq.~(\ref{eq:mh2loop-Ap}) noted above). We observe that the parameter region $A_t \sim m_{U_3}$ yields the correct Higgs mass, while simultaneously minimizing the $\Delta \rho$ parameter. 

\subsection{Stop contributions to Higgs couplings}

In the region of stop parameter space consistent with the observed Higgs mass and a small $\Delta \rho$ parameter, the coupling of the Higgs to the lightest stop is suppressed. Taking the limit $m_{\tilde t_2} \gtrsim X_t \gg m_{\tilde t_1} \sim m_t \gg m_Z$ in the coupling $\lambda_{h^0 \tilde t_1 \tilde t_1^*}$ in Eq.~(\ref{eq:sfermion-higgs}) becomes
\begin{equation}
\lambda_{h^0 \tilde t_1 \tilde t_1^*} \simeq \frac{\sqrt{2} m_t^2}{v}\left(1 - \frac{X_t^2}{m_{\tilde t_2}^2}  \right),
\end{equation}
where $X_t \equiv A_t - \mu \cot \beta$. As discussed above, the $\rho$ parameter is minimized in the regime $X_t \sim m_{\tilde t_2}$, in which case the coupling 
$\lambda_{h^0 \tilde t_1 \tilde t_1^*}$ is suppressed. Thus, we do not expect large corrections to the loop induced Higgs couplings $h \rightarrow gg,\gamma\gamma$ from stop loops. 

\section{Constraints on the Sbottom Parameter Space}\label{sec:combined}

In this section we combine the constraints from precision electroweak data and the Higgs signal strength measurements, identifying the allowed parameter space for light sbottoms. Our aim here is to present conservative constraints, and as such we will assume that the lightest sbottom $\tilde b_1$ is counted as a $b$ quark in the analyses concerning the precision electroweak observables. As can be seen from comparing Figs.~\ref{fig:ewpd-inv},\ref{fig:ewpd-had},\ref{fig:ewpd-b}, the electroweak constraints are the weakest in this case. If instead the sbottom is counted as hadrons or is ``invisible'', the constraints will be considerably stronger than those presented in this section. 

\begin{figure*}
\begin{center}
\includegraphics[width=0.47\textwidth]{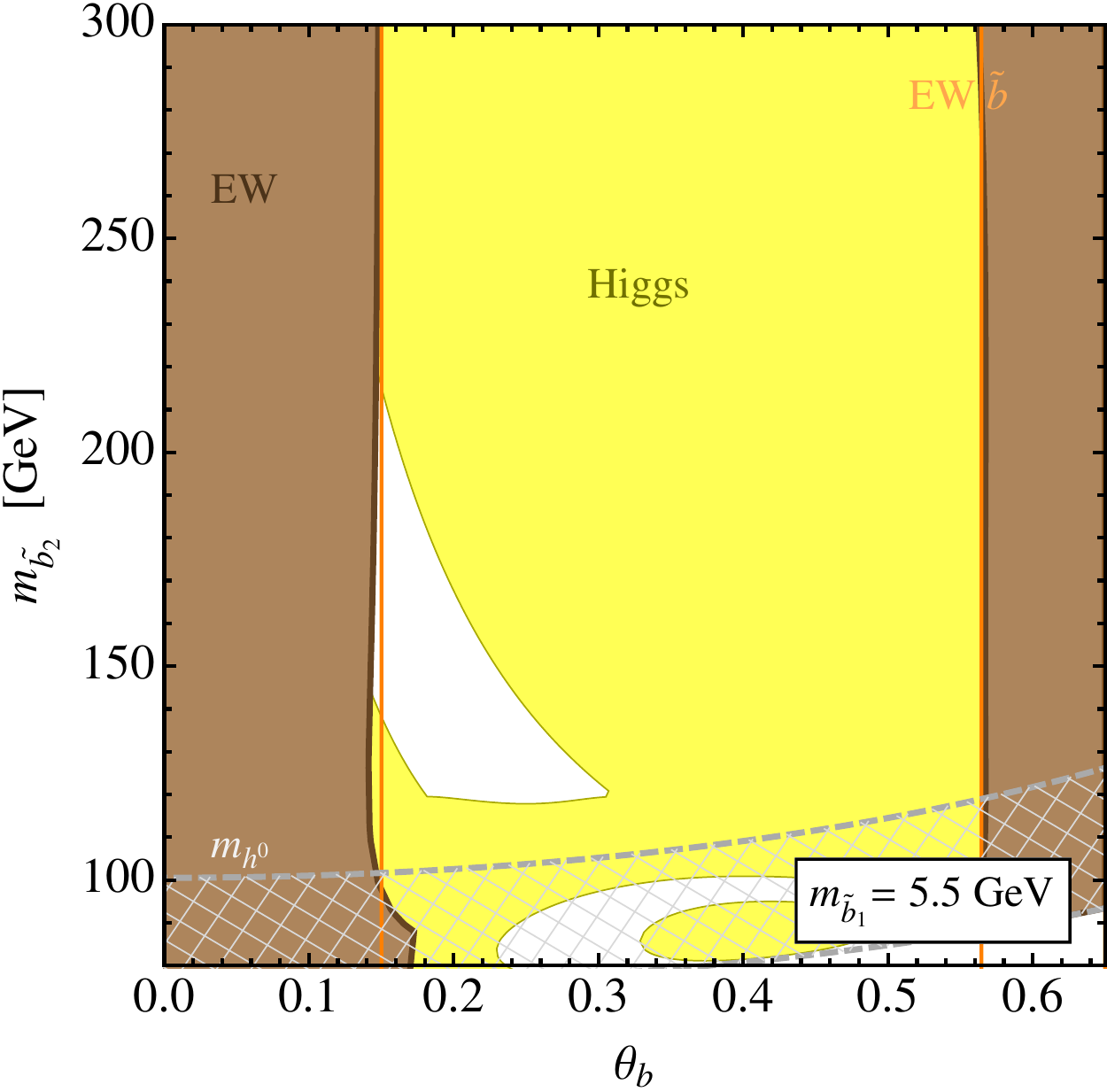}~~~
\includegraphics[width=0.47\textwidth]{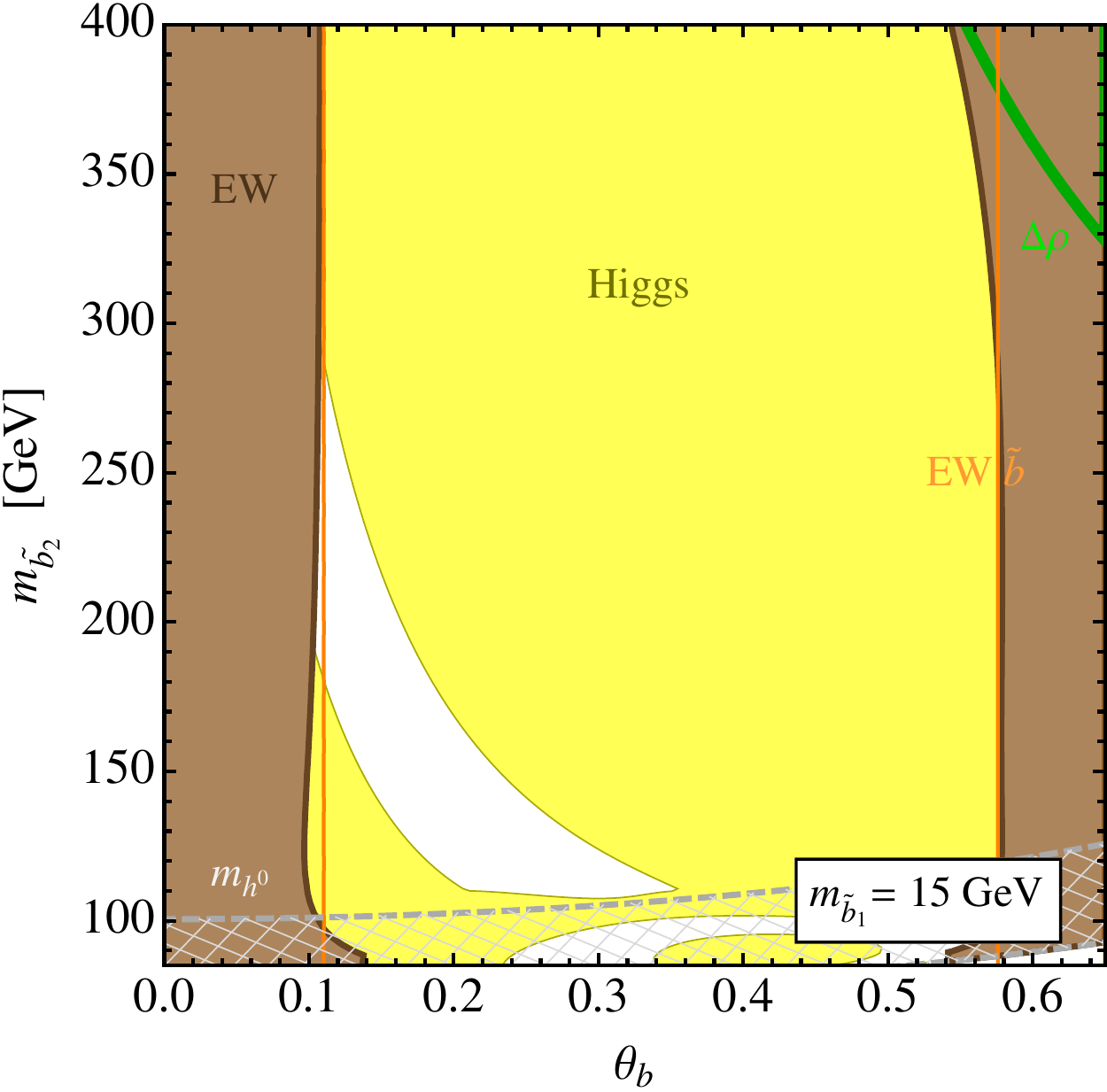}~~~\\
\includegraphics[width=0.47\textwidth]{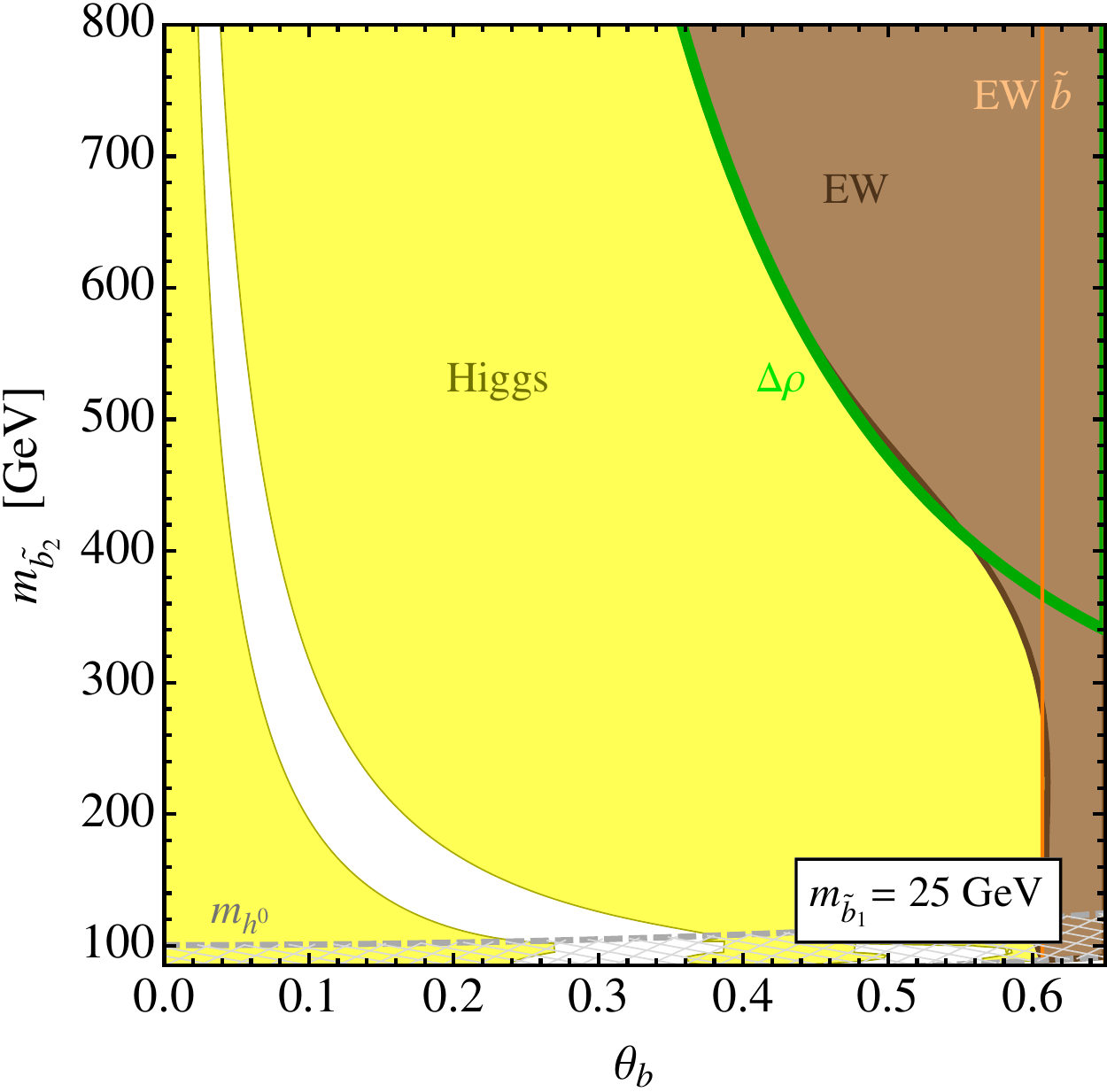}~~~
\includegraphics[width=0.47\textwidth]{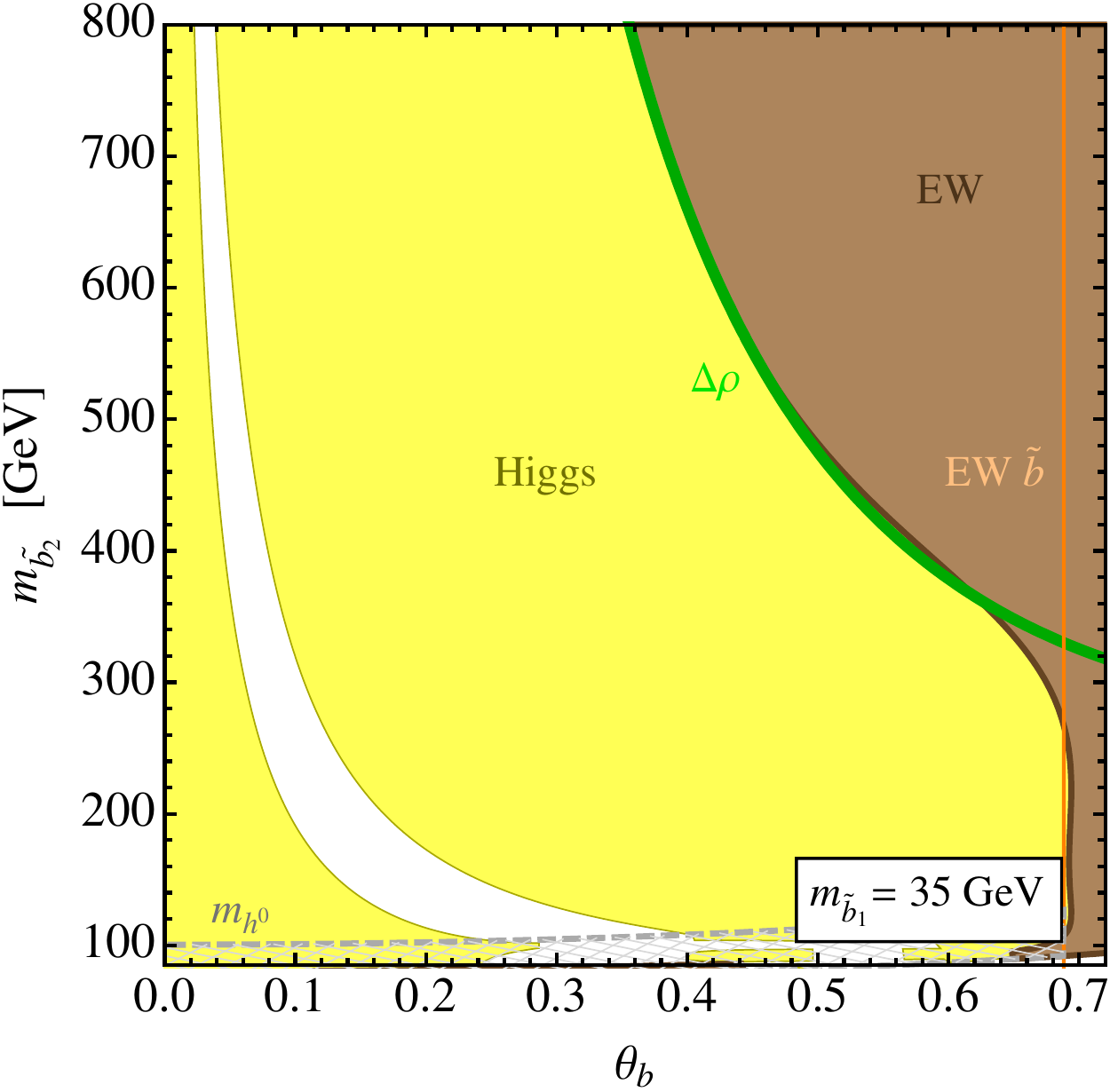}~~~
\caption{ 
\small  
{\it Sbottom parameter space}: Here we represent in white the parameter space in the $\theta_b - m_{\tilde b_2}$ that is allowed by the combined precision electroweak and Higgs signal strength data. For comparison, we also display the excluded regions derived by considering 1) {\it only} the Higgs signal strength data (yellow),
2) {\it only} the precision electroweak data (brown), 3) {\it only} the precision constraints from sbottom decays (orange), and 4) {\it only} the $\Delta \rho$ parameter (green). Finally, in the grey hatched region the predicted Higgs mass is too small within the context of the MSSM. The results are presented for $m_{\tilde b_1} = 5.5, 15, 25, 35$ GeV while we have fixed $\mu = 200$ GeV, $\tan\beta = 10$, $m_{U_3} =A_t = 2$ TeV.}
\label{fig:Combined-1}
\end{center}
\end{figure*}

In Fig.~\ref{fig:Combined-1}, we display the constraints in the $\theta_b - m_{\tilde b_2}$ plane for four cases of the lightest sbottom mass, $m_{\tilde b_1} = 5.5, 15, 25, 35$ GeV. In these plots, we have also fixed $\mu = 200$ GeV, $\tan\beta = 10$, $m_{U_3} = 2.5$ TeV, and $A_t = 2$ TeV. The allowed region from the combined constraints from the Higgs signal strength data and the precision electroweak data is shown in white. In order to understand the shape of this region, we have also displayed the regions allowed by consideration of only a subset of these effects (see the caption of Fig.~\ref{fig:Combined-1} for a detailed explanation). Finally, the gray hatched region indicates where the Higgs mass is too small in the MSSM.

If the first sbottom is very light, Fig.~\ref{fig:Combined-1}, suggests that the second sbottom must also be quite 
light to be compatible with both electroweak and Higgs data. In particular, for $m_{\tilde b_1} = 5.5$ GeV ($m_{\tilde b_1} = 15$ GeV), 
the combined constraints imply that $m_{\tilde b_2}  \lesssim 230$ GeV ($m_{\tilde b_2} \lesssim 300$ GeV). Mixing angles that are too small ($\theta_b \lesssim 0.1$) or too large ($\theta_b \gtrsim 0.6$) are constrained by the precision data, and in particular the tension in $\sigma_{\rm had}^0$ from the presence of the new decay mode $Z\rightarrow \tilde b_1 \tilde b_1^*$. The Higgs signal strength data, which essentially traces out the curve where the $h^0-\tilde b_1- \tilde b_1^*$ coupling is small, further limits the allowed region. The $\Delta\rho$ parameter poses no additional constraint in this example, but depends sensitively on the parameters in the stop sector. 
In the MSSM, the Higgs mass imposes an additional constraint in this example, excluding regions where the second sbottom is lighter than about 150 GeV, but again is highly sensitive to the stop sector. 

Conversely, if the lightest sbottom is somewhat heavier than 15 GeV, the precision electroweak data no longer poses a constraint near $\theta_b \sim 0$ due to the phase space suppression in the decay $Z \rightarrow \tilde b_1 \tilde b_1^*$ (except possibly due to the $\Delta\rho$ parameter, which depends in detail on the values of $m_{U_3}$ and $A_t$). Therefore, in this regime the second sbottom $\tilde b_2$ can be much heavier, as is clearly seen in Fig.~\ref{fig:Combined-1}. Again we emphasize that these are conservative constraints, and if the sbottom is ``counted'' as hadrons or is invisible the constraints are stronger, particularly near the $\theta_b \sim 0$ region of parameter space.

\newpage

%%%%%%%%%%%%%%%%%
%%%%%%%%%%%%%%%%%
\section{Collider constraints}\label{sec:collider}

We have seen in the previous sections that the precision electroweak data and the Higgs signal strength measurements impose tight constraints on the possibility of a light sbottom with mass less than $m_{Z}/2$. Furthermore, 
while the bounds from the precision electroweak data depend on how the sbottom is reconstructed, and thus its precise decay mode, conservative, robust limits are obtained under the assumption that the sbottom is counted as a $b$-quark. The Higgs signal strength data, on the other hand, is not yet sensitive to the precise decay channel of the sbottom. 

Therefore, the constraints from these data, while indirect, have the advantage of not being tightly wed to the model dependent details of the sbottom decays. The main conclusions from this analysis can be inferred from Fig.~\ref{fig:Combined-1}: 1) for $m_{\tilde b_1}  \lesssim 15$ GeV, the second sbottom must be fairly light, $m_{\tilde b_2} \lesssim 300$ GeV, and 2) if the lightest sbottom is somewhat heavier than 15 GeV, then the second sbottom can be made heavy while being consistent with these data. 
In this section, we will discuss the constraints coming from searches at colliders, which provide a more direct probe of the light sbottom scenario, but are strongly dependent on the details of the spectrum and specific decay channels of the sbottom.
We will focus on a canonical scenario with a neutral fermion LSP $\tilde \chi^0$, which could be a bino, gravitino, or singlino, and a sbottom NLSP.
In particular, we will emphasize the importance of the direct constraints on the second sbottom, which, as just mentioned, is predicted to be light if $m_{\tilde b_1} \lesssim 15$ GeV. 

In the scenario under consideration, the lightest sbottom will decay via
\begin{equation}
\tilde b_1 \rightarrow b + \tilde \chi^0.
\label{eq:decay}
\end{equation}
Direct searches for superpartners at $e^+ e^-$ colliders, such as TRISTAN~\cite{Adachi:1988fk}, and LEP~\cite{Heister:2002hp,Abdallah:2003xe,Achard:2003ge,Abbiendi:2002mp}, rule out such light sbottoms $\tilde b_1$ unless a degeneracy in the spectrum exists, in which case some of the decay products will be soft. Let us therefore consider these cases in some detail. 
  
\subsection{Compressed regime}

In the compressed regime, $m_{\tilde b_1} \gtrsim m_{\tilde \chi^0} \gg m_b$, the  LSP carries most of the momentum of the sbottom while the $b$ quark is fairly soft. However, $\tilde b_1 \tilde b_1^*$ pairs will, in the absence of initial state radiation (ISR), be produced back-to-back, and therefore the missing momentum in such events will be suppressed since the two LSPs emitted in the decay also travel back-to-back. Because of this, standard sbottom searches have low efficiencies in this kinematic regime. To get a handle on this region, one can take advantage of events with hard ISR. In such events the sbottoms pairs are boosted and the LSPs are therefore misaligned, leading to a significant missing momentum. For example, Ref.~\cite{Arbey:2012bp} investigated the constraints from the CMS monojet search~\cite{CMS-PAS-EXO-12-048}, concluding that sbottoms heavier than about 24 GeV are ruled out. Furthermore, the LHC sbottom searches~\cite{Aad:2013ija,Chatrchyan:2013lya,Aad:2011cw,Chatrchyan:2012wa}
looking for 2 $b$-jets and missing transverse energy often consider signal regions with an additional hard jet to attack the compressed regime and could be sensitive to such a light sbottom, particularly because the production rate is enormous.  We are aware of a forthcoming study considering the limits from the LHC sbottom searches on very light sbottom pairs produced in association with a hard jet~\cite{liu}. 

Besides direct production of the lightest sbottom, one can look for signatures of the second sbottom $\tilde b_2$, which as emphasized already, must be lighter than about 300 GeV if the first sbottom is below 15 GeV. For instance, the ATLAS sbottom search~\cite{Aad:2013ija} places a limit $m_{\tilde b} > 650\,{\rm  GeV}$, 
under the assumption of a $100\%$ branching ratio for the decay $\tilde b \rightarrow b \tilde \chi^0$. 
In the presence of a very light first sbottom $\tilde b_1$, there will be additional decay modes $\tilde b_2 \rightarrow Z \tilde b_1$, $\tilde b_2 \rightarrow h^0 \tilde b_1$. This will lead to an $O(1)$ suppression in the branching ratio $\tilde b_2 \rightarrow b \tilde \chi^0$ which, however, will generally not be sufficient to evade the constraints on a 300 GeV sbottom.  
One way to evade the constraints on $\tilde b_2$ pair production from these searches would be to consider a gravitino or singlino LSP, since the decay rate $\tilde b_2 \rightarrow b \tilde  \chi^0$ would be suppressed in such a case.  However, in this case, one should consider alternative channels involving the $Z$ boson in the $\tilde b_2$ decay. 
In particular, the signature may show up in SUSY searches with leptons or searches for heavy $B'$ quarks (which decay via $B'\rightarrow Z b$). We will return to the constraints from $B'$ searches momentarily. 

Looking forward, it would be useful for ATLAS and CMS to develop searches with the aim of explicitly probing the lightest sbottom, perhaps by taking advantage of events with hard ISR. It would also be desirable to search for the second sbottom decaying via $\tilde b_2 \rightarrow Z \tilde b_1$ which are relevant for a gravitino or singlino LSP. Such searches could exploit the dilepton pairs from the $Z$, in addition to the $b$-jets and missing transverse energy.

\subsection{Stealth regime}

Besides the compressed regime, another kinematic region which is not covered by sbottom searches at $e^+ e^-$ colliders is the stealth regime, 
$m_{\tilde b_1} \gtrsim m_{b} \gg m_{\tilde \chi}^0$, so named following Refs.~\cite{Fan:2011yu,Csaki:2012fh,Fan:2012jf}. 
In this regime, the LSP emitted in the decay of the sbottom is very soft, and the sbottom is essentially indistinguishable from a $b$ quark. Note that this provides one concrete mechanism by which a sbottom would be counted as a $b$ quark in the precision electroweak measurements. 

First, we note that a very light bino in this case is ruled out by searches for invisible decays of $\Upsilon$(1S). Sbottom exchange 
will mediate the annihilation decay $\Upsilon \rightarrow \tilde B \tilde B$. 
Using the results of Ref.~\cite{Dreiner:2009er} we obtain the branching ratio 
${\rm Br}_{\Upsilon\rightarrow \tilde B \tilde B} 
\approx    3.5 \times 10^{-3}\, (1-\tfrac{3}{2} \sin^2 \theta_b)^2$.
A search from BaBar~\cite{Aubert:2009ae} in events containing $\Upsilon(3S) \rightarrow \pi^+ \pi^- \Upsilon(1S) $ leads to the constraint
${\rm Br}(\Upsilon(1S) \rightarrow ~ {\rm invisible})   < 3.0 \times 10^{-4}.$ 
In order to evade the bound, a mixing angle of $\theta_b \gtrsim 0.76$ is needed, which is well outside the region allowed by the precision electroweak data 
(see Fig.~\ref{fig:Combined-1} for $m_{\tilde b_1} = 5.5$ GeV). Thus, a pure bino LSP is ruled out in this scenario. In principle, it may be possible to weaken this bound  by allowing a large admixture of Higgsino in the lightest neutralino eigenstate, but then one must contend with other challenges, such as new $Z$ and Higgs decays to neutralino pairs and the expectation of very light charged states.  

Instead, one can consider a gravitino or singlino LSP. In this case, we expect that the second sbottom will preferentially decay via $\tilde b_2 \rightarrow Z \tilde b_1$, $\tilde b_2 \rightarrow h^0 \tilde b_1$ rather than directly to the LSP. Therefore, since the $\tilde b_1$ will be reconstructed as a $b$ quark, the signature of second sbottom $\tilde b_2$ will be very similar to that of a heavy $B'$ quark which decays via $B'\rightarrow Z  b$, $\tilde b_2 \rightarrow h^0 b$. 
Searches for $B'\rightarrow Z b$ have been carried out by ATLAS~\cite{ATLAS-CONF-2013-056,Aad:2012pga}, CMS \cite{CMS-PAS-B2G-13-003,CMS-PAS-B2G-12-021,CMS-PAS-EXO-11-066}, and CDF~\cite{Aaltonen:2007je}.  While the CDF~\cite{Aaltonen:2007je} search does not constrain the second sbottom, the ATLAS 7 TeV search~\cite{Aad:2012pga}  explicitly covers masses heavier than 200 GeV, and the quoted upper limit on the cross section appears to rule out the second sbottom in the mass range of 200-230 GeV. Furthermore, the ATLAS 8 TeV search~\cite{ATLAS-CONF-2013-056} explicitly covers masses heavier than 350 GeV, and rules out sbottoms in the mass range 450-450 GeV. The CMS searches explicitly cover masses above 350 GeV but may also have sensitivity to lighter sbottoms. Naive extrapolation of these limits to lower $\tilde b_2$ masses would seem to rule out additional parameter space, although a more detailed study is needed to draw a firm conclusion. The collaborations should carry out searches to explicitly cover the possibility of a second sbottom which mimics a $B'$ quark, but has a lower production rate.  Furthermore, since the lightest sbottom is mostly left-handed, there is also a light stop in the spectrum, which will dominantly decay via $\tilde t_1 \rightarrow W \tilde b_1$. Therefore, the stop will mimic in all respects a fermionic top partner, $T'$, but with a much lower production rate.

\subsection{Other possibilities}

We have given a cursory discussion of the collider constraints on the very light sbottom scenario in the case that the LSP is neutral fermion. The scenario appears to face strong constraints, particularly if $m_{\tilde b_1} \lesssim 15$ GeV, since in this case the second sbottom should be lighter than about 300 GeV and is constrained by various searches at colliders. However, a detailed study and recasting of existing searches should be performed to see if any allowed regions exist. 

One possible way out of these limits is to consider additional light states in the spectrum, {\it e.g.}, electroweakinos, which may open up new decay modes for the second sbottom.
Additionally, one could consider the lightest sbottom to be the LSP and decay through a small R-parity violating coupling. For example, if the sbottom decays through a $UDD$ operator to a pair of jets, then the direct collider constraints on the scenario will be quite weak due to the huge QCD background.  

%%%%%%%%%%%%%%%%%%%%%%%%%%%%%%%%
%%%%%%%%%%%%%%%%%%%%%%%%%%%%%%%%
\section{Discussion and Conclusions}\label{sec:conclusions}

In this paper we have investigated a scenario containing a very light sbottom with mass $m_{\tilde b_1} \lesssim m_Z/2$. We have focused on the indirect, but less model dependent constraints obtained from the sbottom contributions to the precision electroweak and Higgs signal strength measurements. Particularly for a light first sbottom with mass below about 15 GeV, these combined datasets leave open a small region of parameter space and predict that the second sbottom is below about 300 GeV.  However, if the first sbottom is a bit heavier, then the mass of the second sbottom can be raised to much higher values.

The conclusions above are drawn under the assumption that the sbottom is ``counted'' as a $b$-quark. In this case the precision constraints coming from the production of sbottoms through the new decay $Z\rightarrow \tilde b_1 \tilde b_1^*$ and the continuum process  $e^+ e^-\rightarrow \tilde b_1 \tilde b_1^*$ are the weakest. However, the sbottom may be counted as hadrons or alternatively may be ``invisible'', {i.e.}, not populate the signal regions in the searches entering the precision measurements. If so, the constraints from the precision data are much stronger. We emphasize that the measurement of the total hadronic cross section at the $Z$ peak, $\sigma_{\rm had}^0$, plays an important role in constraining a light sbottom and provides a much stronger constraint than the total $Z$ boson width.  This provides a new constraint on the 
dark matter scenario of Ref.~\cite{Arbey:2012bp}, and in particular disfavors their benchmark models.

With the discovery of the Higgs boson, the viable parameter space for a light sbottom has been whittled down to a small region in which the coupling of the Higgs to the first sbottoms is suppressed. With the current precision in the $h\rightarrow b\bar b$ signal strength measurements, these constraints are not sensitive to the manner in which the sbottom decays. Rather, it is the dilution of the branching ratio in the $\gamma\gamma$, $ZZ$, and $WW$ channels that is the primary source of the constraints. 

These constraints are complementary to those provided by direct searches at colliders, since the latter are more strongly dependent on the assumed decay mode of the sbottom. 
We have presented a qualitative overview of the collider constraints for a canonical scenario with a bino, gravitino, or singlino LSP, with minimal assumptions about the spectrum of the lightest states. Dedicated searches should be carried out by ATLAS and CMS to cover the compressed and stealth kinematic regimes in this scenario.

\section*{Acknowledgements}
We are grateful to Patrick Janot for helpful
discussions. 
Work at ANL is supported in part by the U.S. Department of Energy under Contract
No. DE-AC02-06CH11357. L.T.W. and B.B. are supported by the NSF under grant PHY-0756966 and the DOE Early Career Award under grant DE-SC0003930. This work was supported in part by the National Science Foundation under Grant No. PHYS-1066293 and the hospitality of the Aspen Center for Physics.

\appendix
%%%%%%%%%%%%%%%%%%%%%%%%%%%%%%%%%%%%%%%%%%%%%%%%%%%%%%%%%%%%%%%
\section{$h\rightarrow gg, \gamma\gamma$ corrections} \label{appendix}
%%%%%%%%%%%%%%%%%%%%%%%%%%%%%%%%%%%%%%%%%%%%%%%%%%%%%%%%%%%%%%%

One loop sbottom and stop exchange lead to the modification of the $h\rightarrow gg, \gamma\gamma$ rates, which we express in terms of the ratios 
$r_{g,\gamma}  = \Gamma(h^0 \rightarrow gg, \gamma\gamma)/ \Gamma(h^0 \rightarrow gg, \gamma\gamma)_{\rm SM}$. The expressions for $r_{g,\gamma}$ are
\begin{eqnarray}
\label{eq:rg}
r_{g} &=& 
\Bigg\vert  \,
1 \, +
 \, \frac{\sqrt{2}\, v \, C(r)}{A_{gg}^{\rm SM}} \,
  \bigg[  
  \lambda_{h^0 \tilde b_1 \tilde b^*_1}  \frac{A_0(m_{\tilde b_1})}{ m_{\tilde b_1}^2  } + 
   \lambda_{h^0 \tilde b_2 \tilde b^*_2}  \frac{A_0(m_{\tilde b_2})}{ m_{\tilde b_2}^2  }  \\
&&~~~~~~~~~~~~~~~~~~~~~~~~~~~~~~~~~~~~~~~~~~\,  + 
    \lambda_{h^0 \tilde t_1 \tilde t^*_1}  \frac{A_0(m_{\tilde t_1})}{ m_{\tilde t_1}^2  } + 
    \lambda_{h^0 \tilde t_2 \tilde t^*_2}  \frac{A_0(m_{\tilde t_2})}{ m_{\tilde t_2}^2  } 
   \bigg]
\Bigg\vert^2,   \nonumber \\
\label{eq:rga}
r_{\gamma}  &=&   
\Bigg\vert  \,
1 \, - \,
 \frac{  v  \, d(r) }{\sqrt{2} A_{\gamma \gamma }^{\rm SM}}  \,
  \bigg[ Q_{\tilde b}^2 \left(  
   \lambda_{h^0 \tilde b_1 \tilde b^*_1}  \frac{A_0(m_{\tilde b_1})}{ m_{\tilde b_1}^2  } + 
    \lambda_{h^0 \tilde b_2 \tilde b^*_2}  \frac{A_0(m_{\tilde b_2})}{ m_{\tilde b_2}^2  } \right)  \\
    & & ~~~~~~~~~~~~~~~~~~~~~~~~~~~~~~~~~ \,+  
    Q_{\tilde t}^2 \left(  
   \lambda_{h^0 \tilde t_1 \tilde t^*_1}  \frac{A_0(m_{\tilde t_1})}{ m_{\tilde t_1}^2  } + 
    \lambda_{h^0 \tilde t_2 \tilde t^*_2}  \frac{A_0(m_{\tilde t_2})}{ m_{\tilde t_2}^2  } \right)
   \bigg]
\Bigg\vert^2,  \nonumber 
\end{eqnarray}
where $C(r)=1/2$, $d(r)=3$, $Q_{\tilde b}=-1/3$, $Q_{\tilde t}=2/3$, $v=174$ GeV, and 
$A_{gg}^{\rm SM} \approx 1.3$, $A_{gg}^{\rm SM} \approx 6.6$.
The scalar loop function $A_0(m)$ is defined in Ref.~\cite{Djouadi:2005gj} and
approaches  $A_0(m)\rightarrow 1/3$ in the limit $m \gg m_{h^0}$.

In the decoupling limit, the couplings of the Higgs boson $h^0$ to the stops and sbottoms  are given by 
\begin{eqnarray}
\label{eq:sfermion-higgs}
\lambda_{h^0 \tilde b_1 \tilde b_1^*} & = & \sqrt{2} v \left\{  \frac{m_b^2}{v^2}  +\frac{m_Z^2 \cos{2\beta}}{v^2}\left[  s_b^2 \left( -\tfrac{1}{2}+\tfrac{1}{3}s_W^2  \right)
+c_b^2 \left( -\tfrac{1}{3}s_W^2  \right)   \right]  + s_b c_b \frac{m_b (A_b - \mu \tan\beta)}{v^2}    \right\}, ~~~~ \nonumber \\
\lambda_{h^0 \tilde b_2 \tilde b_2^*} & = & \sqrt{2} v \left\{  \frac{m_b^2}{v^2}  +\frac{m_Z^2 \cos{2\beta}}{v^2}\left[  c_b^2 \left( -\tfrac{1}{2}+\tfrac{1}{3}s_W^2  \right)
+s_b^2 \left( -\tfrac{1}{3}s_W^2  \right)   \right]  - s_b c_b \frac{m_b (A_b - \mu \tan\beta)}{v^2}    \right\}, \nonumber \\
\lambda_{h^0 \tilde t_1 \tilde t_1^*} & = & \sqrt{2} v \left\{  \frac{m_t^2}{v^2}  +\frac{m_Z^2 \cos{2\beta}}{v^2}\left[  s_t^2 \left( \tfrac{1}{2}-\tfrac{2}{3}s_W^2  \right)
+c_t^2 \left( \tfrac{2}{3}s_W^2  \right)   \right]  + s_t c_t \frac{m_t (A_t - \mu \cot\beta)}{v^2}    \right\}, \nonumber \\
\lambda_{h^0 \tilde t_2 \tilde t_2^*} & = & \sqrt{2} v \left\{  \frac{m_t^2}{v^2}  +\frac{m_Z^2 \cos{2\beta}}{v^2}\left[  c_t^2 \left( \tfrac{1}{2}-\tfrac{2}{3}s_W^2  \right)
+s_t^2 \left( \tfrac{2}{3}s_W^2  \right)   \right]  - s_t c_t \frac{m_t (A_t - \mu \cot\beta)}{v^2}    \right\}.
\end{eqnarray}

\end{document}